\documentclass[aps,prb,twocolumn,unsortedaddress,floatfix,nofootinbib]{revtex4-2}
\usepackage{multirow}
\usepackage{amsthm}
\usepackage{amsfonts}
\usepackage{siunitx}
\usepackage{amsmath}
\usepackage{amssymb}
\usepackage{graphicx}
\usepackage{verbatim}
\usepackage[colorlinks]{hyperref}
\usepackage{tikz}
\usepackage{pgfplots}
\usepackage{braket}
\usepackage{xcolor}
\usepackage{physics,siunitx,algorithm,algorithmic,listings,siunitx}
\usepackage[normalem]{ulem}
\usepackage{DejaVuSansMono}

\usepackage{fancyhdr}
\definecolor{dwteal}{RGB}{23,190,187}
\definecolor{dwblue}{RGB}{42,125,225}
\definecolor{dworange}{RGB}{243,120,32}
\definecolor{dwgray}{RGB}{134,134,134}

\definecolor{linkcolor}{RGB}{0,83,166}
\hypersetup{
  colorlinks = true,
  allcolors = {linkcolor}
}

\begin{document}
\title{Hybrid quantum annealing for larger-than-QPU lattice-structured problems}

\author{Jack~Raymond, Radomir~Stevanovic, William~Bernoudy, Kelly~Boothby, Catherine~McGeoch, Andrew~J.~Berkley, Pau~Farr\'e, Andrew D.~King}

\address{D-Wave Systems, Burnaby, BC Canada}

\date{\today}

\begin{abstract} Quantum processing units (QPUs) executing annealing algorithms have shown promise in optimization and simulation applications. Hybrid algorithms are a natural bridge to additional applications of larger scale.  We present a straightforward and effective method for solving larger-than-QPU lattice-structured Ising optimization problems.  Performance is compared against simulated annealing with promising results, and improvement is shown as a function of the generation of D-Wave QPU used.
\end{abstract}

\maketitle

\section{Introduction}

In this report, we use a hybrid quantum annealing approach to search for the ground state of a classical Ising model Hamiltonian, parameterized by couplers $\{J_{ij}\}$ and external fields $\{h_i\}$
\begin{equation}
  \mathrm{argmin}_x H(x)\;; \qquad H(x) = \sum_{i<j} J_{ij} x_i x_j + \sum_i h_i x_i\;, \label{eq1}
\end{equation}
with $x \in \{-1,1\}^N$.
Ising model ground state search is equivalent to quadratic unconstrained binary optimization (QUBO), an NP-complete problem~\cite{Barahona_1982}.

D-Wave QPUs can implement a transverse field quantum annealing algorithm for this task~\cite{Kadowaki:QA,Johnson2011,Boothby2021}.
The Advantage\textsuperscript{TM} QPU can encode Ising optimization problems with over 5000 variables, coupled according to a Pegasus lattice structure (a special connectivity-15 graph)~\cite{Boothby:19},
and problems of diverse graphical structure can be solved by minor embedding~\cite{Choi2011}.
However, many problems of interest cannot be minor embedded on QPUs of practical scale. We examine a simple hybrid optimization method for such cases, exploiting the computational power of QPUs up to the maximum programmable scale. Results are presented for spin-glass models; spin glasses are a well-studied exemplar of random hard optimization problems~\cite{Nishimori2001,Harris2018}.
For these models, quantum annealing (QA) and simulated (thermal) annealing (SA) are well-motivated heuristics for obtaining good upper bounds on the ground state on short timescales~\cite{Kadowaki:QA,Harris2018,Kirkpatrick:OSA,doi:10.1126/science.284.5415.779}; we examine performance of these methods alongside other canonical algorithms.

QA in the Advantage QPU operates with flux qubit states, which can be approximated as Ising spins~\cite{Johnson2011,Boothby:19}. 
A system Hamiltonian $H$ is prepared, initially dominated by a driver term $H_D=-\sum_{i \in V} \sigma^x_i$, so that spins are prepared in a uniform superposition, where $V$ indexes the programmable qubits and $\sigma_i$ are Pauli operators.
The Hamiltonian is evolved smoothly towards a problem Hamiltonian
\begin{equation}
  H_P = \sum_{ij \in E} J_{ij}\sigma^z_i \sigma^z_j + \sum_{i \in V} h_{i}\sigma^z_i\;,
\end{equation}
where $E$ indexes the programmable couplers.\footnote{If the target Hamiltonian matches the QPU architecture, the programmed couplers can be chosen equal to the target Hamiltonian (\ref{eq1}). In other cases minor embedding determines a transformation between the target model and programmed model~\cite{Choi2011}.}
The adiabatic theorem guarantees that slow evolution achieves a ground state of the problem Hamiltonian~\cite{Farhi2001}.
At the end of the anneal, states decohere into the computational (flux) basis where they are measured, and can be mapped to solutions of the optimization problem.
In practice, evolution is subject to a finite evolution time, analog noise and model non-idealities, so that we arrive at a sample drawn from a distribution over low energy states, without a guarantee of optimality~\cite{Harris2008,Amin2009a}.
If this distribution has significant probability measure on the ground state, we can repeat the anneal multiple times to obtain an optimal solution with confidence, otherwise obtaining a low-energy state.

One of the simplest algorithms available to exploit a powerful subsolver is large neighborhood local search (LNLS).
LNLS has been the basis for many quantum hybrid methods, and classical alternatives, such as D-Wave's qbsolv utility and the Hamze-de~Freitas-Selby algorithm~\cite{Booth2017,hamze2004fields,SelbyGITHUB}. 
An application particularly close to that presented in this work was considered by Okada et al.\ in 2019~\cite{Okada2019}:
as in this paper, they considered an LNLS algorithm, with a QPU subsolver, applied to lattice-structured spin glasses.
The main difference in this paper, aside from developments in the QPU used, is that we consider the use of embeddings that are optimized a priori for the specific task of lattice solving, whereas in that paper lattices were used as an exemplar for testing performance of a more general embedding heuristic within an LNLS workflow.

In Section \ref{AlgOut} we introduce the spin-glass ensembles we study, the process of minor embedding required for use of the QPU as a lattice subsolver, and the integration of that solver into a basic LNLS heuristic. In Section \ref{results} we show performance for different embedding methods, different QPUs, and a comparison against alternative competitive heuristics.
Appendix \ref{app:problem_definitions} contains further details on the spin-glass ensemble studied and analysis of additional model classes and optimization methods. This includes demonstration of the hybrid method in the context of planted solutions, where attainment of an optimum is verifiable. Details of our SA implementations are discussed in Appendix \ref{app:SA}. Appendix \ref{app:hybriddetails} contains further details on our open-source implementations including basic code, QPU parameterization details, and analysis of alternative hybrid workflows.

In this paper we show that an LNLS-based hybrid method enabled by QPUs efficiently samples low-energy states, with almost all the work being done by the QPU.
We demonstrate the effectiveness of lattice-specific embeddings relative to heuristic and generic-graph options.
The D-Wave Advantage  QPU is shown to outperform the D-Wave 2000Q\textsuperscript{TM} QPU; we can attribute the success to size and embedding efficiency.
The hybrid QA method is shown to be competitive with SA applied directly to the full problem, and superior to simple greedy methods.
Our algorithm is implemented in open-source software, allowing reproduction of results.

\section{Algorithm outline}
\label{AlgOut}
\subsection{Lattice and problem definitions}

We consider heuristics applied to periodic simple cubic, and toric-Pegasus, lattice structured problems. Periodicity simplifies, for didactic purposes, the analysis presented. A periodic cubic lattice at scale $L$ contains $L^3$ variables each of connectivity 6. A toric-Pegasus graph at scale $L$ contains $24 \times L^2$ variables each of connectivity 15. Further details are contained in Appendix \ref{app:periodic_lattices}. We expect our method to generalize well to other lattices (or lattice subgraphs). 

An exemplar for hard optimization problems over lattices are spin glasses~\cite{Baity-Jesi2017}. For a given lattice defined by a vertex and edge set $\{V,E\}$, a standard $\pm J$ spin-glass instance can be created by setting couplers $\{J_{ij} : ij \in E\}$ to random values independently and uniformly distributed on $\{-1,1\}$, and setting external fields to zero $\{h_{i}=0 : i \in V\}$. For this random ensemble we are interested in a typical case: the average performance with respect to a collection of sampled instances. In Appendix \ref{app:spin_glasses} we further elaborate upon the useful properties of spin glasses for a study of this nature, while in Appendix \ref{app:planted} we present analyses for some other model ensembles, including simple models with planted solutions where verifying attainment of the optima, or the energy gap to optimality, is simplified.

Experiments discussed in this section involve lattices defined with approximately twice as many variables as can be solved by a conventional minor-embedding process on the QPU as shown in Table \ref{table:minorembedding}. For the cubic lattice case we seek to solve by an LNLS hybrid method $L\times L \times L$ periodic lattice problems. With Advantage QPUs as the subsolver, we target a scale $L=18$ ($N=5832$), and for D-Wave 2000Q QPUs subsolvers, a scale of $L=10$ ($N=1000$). For the case of toric-Pegasus problems we use Advantage QPUs as the subsolver, and target an $L \times L \times 24$ variable problem with $L=22$ ($N=11616$).

\subsection{Minor embedding}
\label{embedding}
Minor embedding allows Ising problems that do not match the QPU architecture to be solved by annealing on D-Wave 2000Q and Advantage QPUs~\cite{Choi2011}.  Multiple coupled qubits (chains) are used to represent one problem variable. Long chain length is understood to hinder performance; among other factors, we can attribute weaker performance to the longer time-scales associated to tunneling of multiple qubits. Determining an embedding using chains of minimal length is NP-hard, and heuristics must be employed~\cite{Cai2014,Okada2019,ocean-SDK}. Naively, it might be assumed that for every call to the QPU we must employ a potentially expensive minor-embedding process on the fly, or else risk paying a large performance penalty from use of a poorly chosen embedding.

Various strategies can be adopted to work around this; one is use of a fully-connected graph (also called a clique) embedding. Any problem defined over $N$ or fewer variables can exploit a clique embedding once it is known, since all lattices are subgraphs of cliques.
Thus we can generate such an embedding once and reuse it. For QPU architectures the most efficient forms of embedding are well understood for cliques~\cite{Boothby:19,cliquepaper,Zephyr}. Enumerating a small number of these, or generating them as needed can be efficient. Thus a simple option is to employ an algorithm exploiting such embeddings, bypassing the need for an expensive embedding algorithm.
Gains in embedding efficiency have been made for various graph types, including cliques, between D-Wave QPU generations. Nevertheless, clique sizes are relatively limited --- embedding efficiency (few  variables per programmable qubit) and performance is sacrificed for reusability.

As demonstrated by Okada et al., this clique-embedding strategy proves to be a poor choice in the context of LNLS for lattices~\cite{Okada2019}. It is possible to create on-the-fly embeddings efficiently that outperform clique embeddings. However, heuristic embeddings obtained this way will typically underperform an optimal embedding in the context of large lattices.\footnote{An embedding that minimizes time to solution under annealing might be considered optimal, and anything else suboptimal. This criteria is in practice substituted for a tractable (yet strongly correlated) criterion, such as the requirement that the largest chain length be minimized. For the lattices we consider embeddings meet this latter criterion~\cite{Choi2011,Boothby:19}.}

The problem sequences generated in the case of LNLS as outlined in Section \ref{LNLS} align with subspaces of the full lattice being studied.
Lattice-structured problems have the appealing quality that subspaces have predictable structure. For example, an $8\times 8 \times 8$ subspace of a cubic lattice is itself an $8\times 8\times 8$ cubic lattice, regardless of its location in the larger problem. The subproblems that are solved in sequence by the subsolver can be chosen to have a common lattice structure. As such we can generate one (or a small number) of good embeddings, and reuse these throughout an LNLS algorithm, for many different subproblems.

Lattice embeddings are not unique, but in many cases one can determine near-optimal, or provably-optimal, embeddings with modest chain length and many variables. The cubic lattice is a favorable case compared to cliques, because we only require two qubits per chain on Advantage QPUs, and the embedding has a regular structure \cite{King2020}. For subproblems derived from our toric-Pegasus lattices the situation is even more favorable, as we have a one-to-one mapping of variables to qubits on the Advantage QPU. Table \ref{table:minorembedding} shows the scale of problem embeddable for various lattices for D-Wave 2000Q and Advantage QPUs. Unless stated otherwise, the hybrid methods enabled by the QPU use the largest available embeddings ($N\lesssim 2700$ for cubic, $N\lesssim 5400$ for Pegasus).\footnote{The subsolver used for Pegasus spin glasses is restricted to the $15\times15\times24$ {\it nice} Pegasus subgraph of $P[m=16]$; the set of more highly connected variables near the center of the QPU~\cite{docs,ocean-SDK}.} Fabrication imperfections in QPUs modify slightly the number of operational qubits and couplers in the working graph (the yield). Consequences of incomplete yield are discussed in Appendix \ref{app:vacancies}.

\begin{table}[t!]
  \begin{center}
    \caption{\label{table:minorembedding} Properties of optimal embeddings by lattice class and QPU solver at maximum embeddable scale, given a fully-yielded QPU~\cite{Boothby:19,King2020,cliquepaper}. $N$ is the maximum size of the graph minor achievable under particular regular embedding procedures, and $C_L$ is the associated chain length. For lattices the structure is also indicated in parentheses, where $C[m]$ and $P[m]$ denote Pegasus and Chimera graphs respectively~\cite{Boothby:19,docs}.}
    \vspace{2mm}
    \scalebox{0.94}{\begin{tabular}{l|l|l|l|l|}
      \textbf{Minor} & \multicolumn{2}{l|}{\textbf{Advantage}, $P[m=16]$} & \multicolumn{2}{l|}{\textbf{D-Wave 2000Q}, $C[m=16]$}\\
          \textbf{} & $N$ & $C_L$ & $N$ & $C_L$\\
      \hline
      Clique & 182 & $\leq$ 17 & 64 & 17\\
      Cubic & 2700 (15x15x12) & 2 & 512 (8x8x8)& 4 \\
      Pegasus & 5640 (P[m=16]) & 1 & 264 (P[m=4]) & $\leq$ 5 
    \end{tabular}}

  \end{center}
\end{table}

\subsection{Large neighborhood local search}
\label{LNLS}

An LNLS strategy for optimization involves iteration of a ground-state estimate for the full problem (\ref{eq1}), which can initially be uniformly random.
On each iteration a subset of variables is selected for construction of a problem at programmable scale.
The subproblem is solved (heuristically or exactly) and the result reintegrated into the global approximation.
In a greedy implementation we can integrate samples such that energy never increases on each update, so that either we arrive at a ground state, or are trapped by a local minimum.
In the latter cases we can restart the algorithm, as resources allow, to improve the ground-state estimate, exploiting randomization in the algorithm to avoid systematic trapping by the same suboptima.
A greedy strategy may not be optimal; accepting non-decreasing energy proposals may allow escape from local minima. Nevertheless, the greedy strategy is simpler for proof of concept and proves sufficient for the demonstrations in this paper. 

The subproblem solved is the conditional energy minimization of a Hamiltonian; conditioned on fixed values for variables outside a subspace $R$.
If the energy can be lowered, reassignment of subproblem variables according to the subproblem optima is then guaranteed to strictly lower the energy of the full problem.\footnote{A heuristic subsolver such as QA may not return an optimum. In such cases the current assignment and proposal from the heuristic can be compared, and the best chosen, so that energy never increases per iteration.}
A problem compatible with the subspace structure must be solved, but with external fields modified by the fixed variable assignments outside the subspace (see Algorithm \ref{pseudo} for details).
If the fields are consistent with a ground state, then the conditioning is optimal and by solving the subproblem we also assign values consistent with a ground state solution.
The algorithm can also correctly assign values to some part of the subspace even if the conditioning values are inaccurate.
If variables at the center of the subspace are weakly correlated with the (distant) values at the boundary,
they may organize into a globally optimal arrangement in spite of poor conditioning.
The premise of the algorithm is that while conditioning values are initially poor, they are iteratively improved.
Unless dynamics are trapped by local minima due to long-range correlations in the way variables must be set,\footnote{In a qualitative sense, we mean longer than the scale of the large neighborhood tackled by the subsolver.}
convergence to the ground state occurs after several updates of every variable. 

Note that an important feature of subspaces for algorithmic success is that there exists within the subspaces variables far from the boundary (the fixed variables),
that are (at least for intermediate stages of the algorithm) weakly correlated with boundary values,
yet strongly correlated with other variables in the subspace.
To achieve this, we choose subspaces that are, to the extent possible, connected and ball-like, with a small surface area relative to volume.
As the subspace volume scales the ability to engineer a favourable surface to volume ratio is limited by the dimensionality of the lattice to which the hybrid method is being applied. Embedding efficiency (variables modeled per qubit programmed) is similarly limited by dimensionality. We anticipate the most favouable outcomes in lower dimensional lattice applications.
In practice, we choose square (at cellular level) subspaces for the toric-Pegasus, and cuboid subspaces for the simple-periodic-cubic lattice.

A complicating factor in the use of QPU solvers is yield: the fraction of operational qubits and couplers. In our implementation, we accommodate incomplete yield by allowing vacancies (internal boundary) in the subspace, to the detriment of the boundary-volume ratio. The consequences are discussed in Appendix \ref{app:vacancies}. If regions are chosen with {\it smooth} boundaries (fewest possible spanning couplers per variable), this is also favorable for QPU solvers, as it constrains the scale of $h$ values that might need to be programmed (see (\ref{eq:subproblem}), Algorithm \ref{pseudo}). A significant development in recent Advantage solvers was the extension of the programmable ``{\small\texttt{h\_range}}'' from $[-2,2]$ to $[-4,4]$, allowing use of larger energy scales.

Selection of the subspaces is an important consideration in LNLS methods.
For simplicity of analysis, we take the approach to choose a subspace uniformly at random on every iteration (maximum programmable-scale sub-cuboids, or Pegasus subgraphs, respectively for our two types of problem). The homogeneous and random nature of spin-glasses makes this a natural starting point for analysis. However, as with the greedy strategy, it should be assumed that we might do better by incorporation of state- and instance-dependent information in the region choice.\footnote{An example of a state and problem-dependent strategy for region selection is the EnergyImpactDecomposer, available in the dwave-hybrid package~\cite{Booth2017,ocean-SDK}.
One can also make use of, as one example, nucleation strategies to propagation information in a guided way between solved regions~\cite{Rams2016}, or conditional independence between regions, for efficiencies.}

Algorithm \ref{pseudo} presents the pseudocode for the method described. A flow diagram is shown in Figure \ref{fig:flow}, using an example of a square lattice of dimensions $8\times8$ with a subsolver of dimensions $4\times4$. A small number of QPU parameters take non-defaulted values in our implementation, as discussed in Appendix \ref{app:parameterization}. The embedding ($M$) is selected uniformly at random from known (precomputed optimal) cases, and the random subspace ($R$) compatible with the embedding is selected uniformly at random. For spin glasses we set $E_{target}=-\infty$ (ground state is unknown) and allow the algorithm to time-out at 128 iterations. In some Appendix examples where ground-state energy is known, we can use this as $E_{target}$.

D-Wave hybrid (dwave-hybrid) is a general, minimal Python framework for building hybrid asynchronous decomposition samplers for Ising optimization problems~\cite{ocean-SDK}.
Given the target Ising model we can implement this algorithm straightforwardly, and vary the workflow to incorporate additional classical methods; variations on the default workflow are considered in Appendix \ref{app:python}. Reference examples are included in dwave-hybrid, and performance of more carefully optimized code is considered in Appendix \ref{app:timing}.

\begin{figure*}
\begin{minipage}{\linewidth}
\begin{algorithm}[H]
    \caption{Minimize $N$ variable lattice problem: $H(x) = \sum_i h_i x_i + \sum_{ij} J_{ij} x_i x_j$ }
    \begin{algorithmic}[1]
      \STATE{Select uniform random $x \in \{-1,1\}^N$}
      \STATE{Create or load embedding(s) compatible with lattice-topology and QPU-architecture}
      \WHILE{$H(x)>E_{Target}$ and time-out not exceeded}
      \STATE{M = random embedding}
      \STATE{R = random subspace compatible with M}
      \STATE{Create subproblem: \begin{equation} H_R(y) = \sum_{ij \in R} J_{ij} y_i y_j + \sum_{i \in R} \left(h_i + \sum_{j \not\in R} J_{ij} x_j\right) y_i \label{eq:subproblem}\end{equation}}
      \STATE{Embed: Create QPU-architecture compatible problem $H_P$, using $M$ and $H_R$}
      \STATE{Program/Readout: Make QPU-API call returning qubit state assignments (samples) $\{x^{(q)}\}$, using $H_P$}
      \STATE{Create variable proposal: Map samples to lattice-subspace compatible values $\{x_R'\}$, using $M$ and $\{x^{(q)}\}$}
      \STATE{Select best proposal: ${\hat x}_R = \mathrm{argmin}_{y \in \{x_R'\}} H_R(y)$}
      \IF{$x_R \leftarrow {\hat x}_R$ lowers $H(x)$}\STATE{$x_R = {\hat x}_R$}
      \ENDIF
      \ENDWHILE
      \RETURN x
    \end{algorithmic}
    \label{pseudo} 
\end{algorithm}
\end{minipage}
\end{figure*}

\begin{figure*}[htb!]
  \begin{center}
    \includegraphics[width=\linewidth]{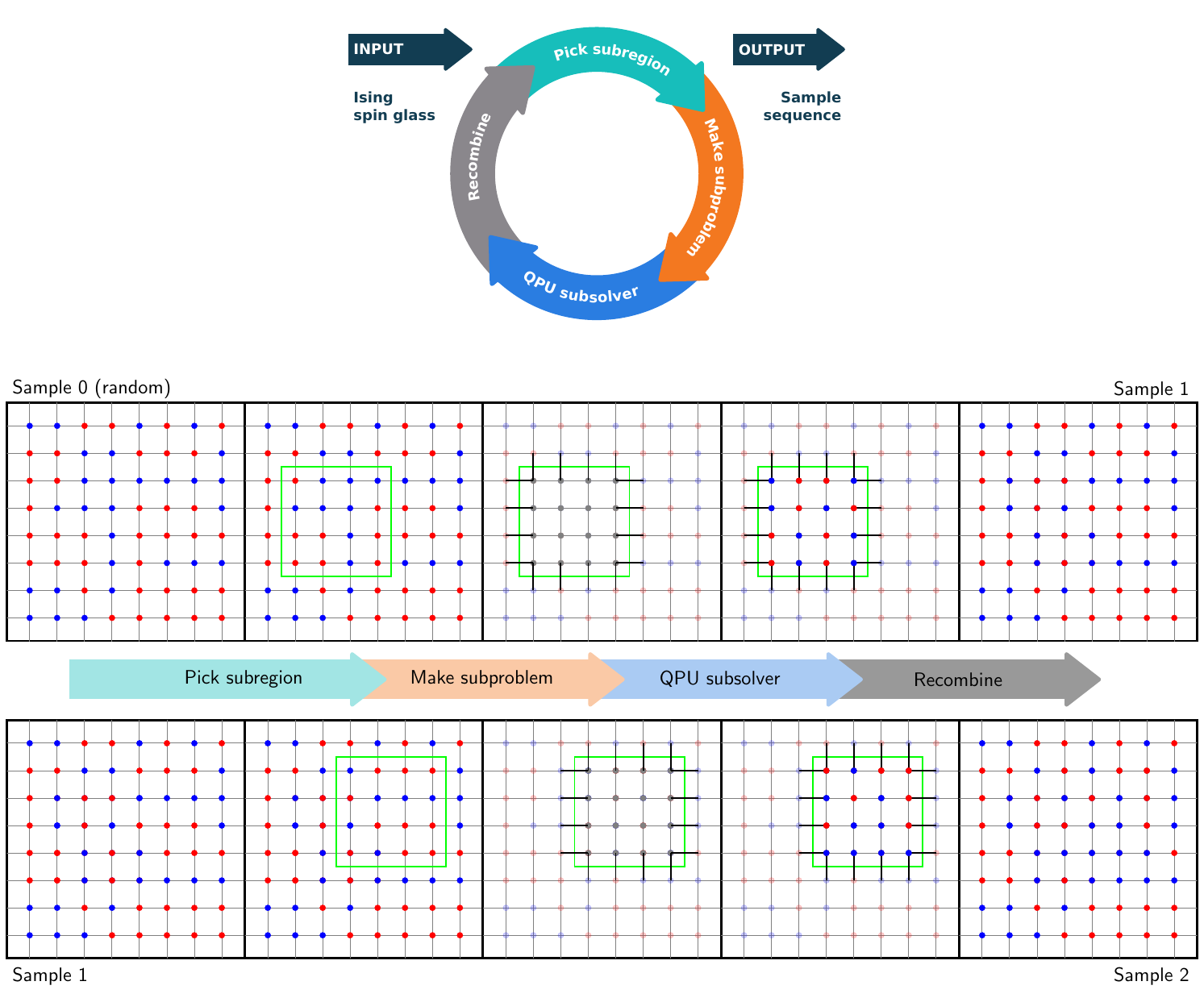}
    \caption{\label{fig:flow} From an initial (random) estimate to the ground state, subspaces are randomly selected and updated. This is iterated to produce heuristic estimates to the ground state of decreasing energy. We restrict analysis in this paper to periodic (simple-cubic or toric-Pegasus) lattices so that there is no boundary --- subspaces consist of regular connected variable sets regardless of displacement.}
  \end{center}
\end{figure*}

\section{Results}
\label{results}

We evaluate performance of our algorithm in terms of burn-down: the energy achieved as a function of time resources used.

Energy can be evaluated at each iteration of our algorithm. As algorithms approach the ground-state energy it is also useful to have a measure resolving the energy gap to the ground state, which can be relatively small compared to instance-to-instance fluctuations in the ground-state energy and energy gaps to optimality early in the algorithm.
As previously noted there is no practical means to firmly establish the ground-state energy for large spin glasses. Nevertheless, we can obtain an upper bound on the ground-state energy by use of a portfolio of solvers. In particular, we run simulated annealing algorithms for long timescales as discussed in Section \ref{app:SA}. We call this estimate $E_0$, and for an optimizer providing a best sample $s$, we can evaluate performance in terms of relative error
\begin{equation}
  r(s) = (E_0 - H(s))/E_0 \;. \label{relativeerror}
\end{equation}
Uniform random samples achieve in expectation a value of 1; samples matching the ground-state estimate achieve a minimal ratio of zero. Relative error is preferred to energy in the majority of plots owing to the efficiency in differentiating heuristics on a range of time and energy scales. Note that if the ground-state energy estimate $E_0$ is larger than the true ground-state energy, then $r(s)$ will be underestimated as a function of the estimator quality. However, the relative ordering of any methods evaluated is not impacted, which is of primary interest. 

Unfortunately, the challenges of time measurement in distributed computation~\cite{10.1145/1007771.55618,10.1145/860575.860711,Drozdowski:02} mean that it is not possible to obtain wall-clock times for the full hybrid computation, in a way that would meet acceptable levels of both precision and replicability. Instead, we report times for the components that can be measured precisely, allowing our results to be reproducible with existing dwave-hybrid code.\footnote{QPU access time is returned by the API, and can be logged.} We provide estimates for additional overhead costs that might affect latency of the full distributed computation, as follows:
\begin{itemize}
\item Algorithm \ref{pseudo} is implemented in dwave-hybrid (Python), and in part as an open-source C++ code. dwave-hybrid is an easily accessible framework for prototyping and benchmarking, but is significantly slower than the C++ implementation. In Appendix \ref{app:timing} we analyze the C++ implementation with the QPU-API call on line 8 substituted by a random sample generator. This indicates that Algorithm \ref{pseudo}---excluding the QPU-API call---requires approximately half a millisecond per iteration, with negligible startup overhead, with some optimization. This software overhead (Algorithm \ref{pseudo}, excluding the QPU-API call) is treated as negligible, and excluded from timing analysis.
\item A portion of the QPU-API call is accounted for by QPU access time~\cite{docs}. The QPU access time is returned by the QPU-API, fluctuating little between iterations, and is proportional to the cost currently charged to a customer.  QPU access time consists of the time spent in programming ($t_p$) and the time spent sampling, the latter being proportional to the number of samples requested ($n_{r}$). Per sample, it accounts for the combination of requested annealing time ($t_a$), and per-readout overheads ($t_{ro}$, primarily thermalization, and the readout process itself). Each iteration invokes the QPU, returning one sample set; the associated latency is accumulated as
  \begin{equation}
    \mathrm{QPU\;access\;time} = \sum_{\textit{sample-sets}}\hspace{-4mm} t_{p} + [t_{ro} + t_{a}] n_{r}\;.\label{eq:QPUaccess}
  \end{equation}
  Using an anneal duration of \SI{100}{\micro s} with 25 reads on the Advantage\_system4.1 solver programming at full scale, QPU access time is approximately \SI{17.5}{\milli s} per iteration.  
\item Additional overhead in the QPU-API call from distributed computation may accrue due to costs of network
  transmission, queueing, and system computations on the server side. Independent
  tests suggest that half a second is a reasonable baseline estimate for round-trip time per
  Advantage QPU invocation using the Ocean SDK sampling clients running from Amazon Web Services\textsuperscript{TM}~\cite{docs,ocean-SDK}.\footnote{Amazon and AWS are trademarks of Amazon Technologies, Inc.} This overhead would impact the
  total computation time of a production-quality hybrid lattice solver to an unknown
  degree, depending on factors such as system traffic and the degree of
  (asynchronous) concurrency achievable for a given input.
  Due to this uncertainty, these overhead costs are not included in reported runtimes.
  It is worth noting that the algorithm presented might be adapted to mitigate for much of this time through workflow changes such as parallelization of job submissions, or in principle by implementation changes such as operation closer to, or on, the QPU-server.
\end{itemize}
In summary, we use QPU access time as a measure of time for our hybrid methods. The end-to-end open-source dwave-hybrid implementation requires significantly longer in practice, but is sufficient to reproduce reported results. The actual time spent by the QPU doing computational work (accumulated annealing time) is significantly shorter, as demonstrated in Appendix \ref{app:tsQPUp}.

The problems studied are large spin glasses. Analysis of time to solution for a variety of algorithms, including programmable scale lattices for QPUs, is well understood~\cite{King2020,KolnServer,McGeoch2013,Coffrin2017,TroyerSTA}. For the lattice types and scales we study, methods that verify optimality require impractical timescales, whereas methods that provide a plausible self-consistent measure of optimally tend also to produce suboptima on the timescales of interest, including the hybrid method we present. We demonstrate some special cases aligned with this intuition in Appendices \ref{app:planted} and \ref{app:SA}.
With this in mind we take the objective to be the attainment of a best upper bound to the ground state, minimizing the energy given some constrained timescale. 

Since we are interested in typical instance performance, we use median energy as our principal measure of algorithmic performance. The algorithm and the target problem class are both random; the median is taken with respect to these random elements. Calculation of the median is further discussed in Appendix \ref{app:python}. Unless stated otherwise, medians are estimated with respect to a common set of 25 random instance realizations (for relative error, the associated $E_0$ estimates are also fixed for all plots). Standard confidence intervals (68\%) are presented for the estimators used in this report.

\subsection{Embedding efficiency and QPU generation}
\label{sec:emb}

\begin{figure}[b]
    \includegraphics[width=\linewidth]{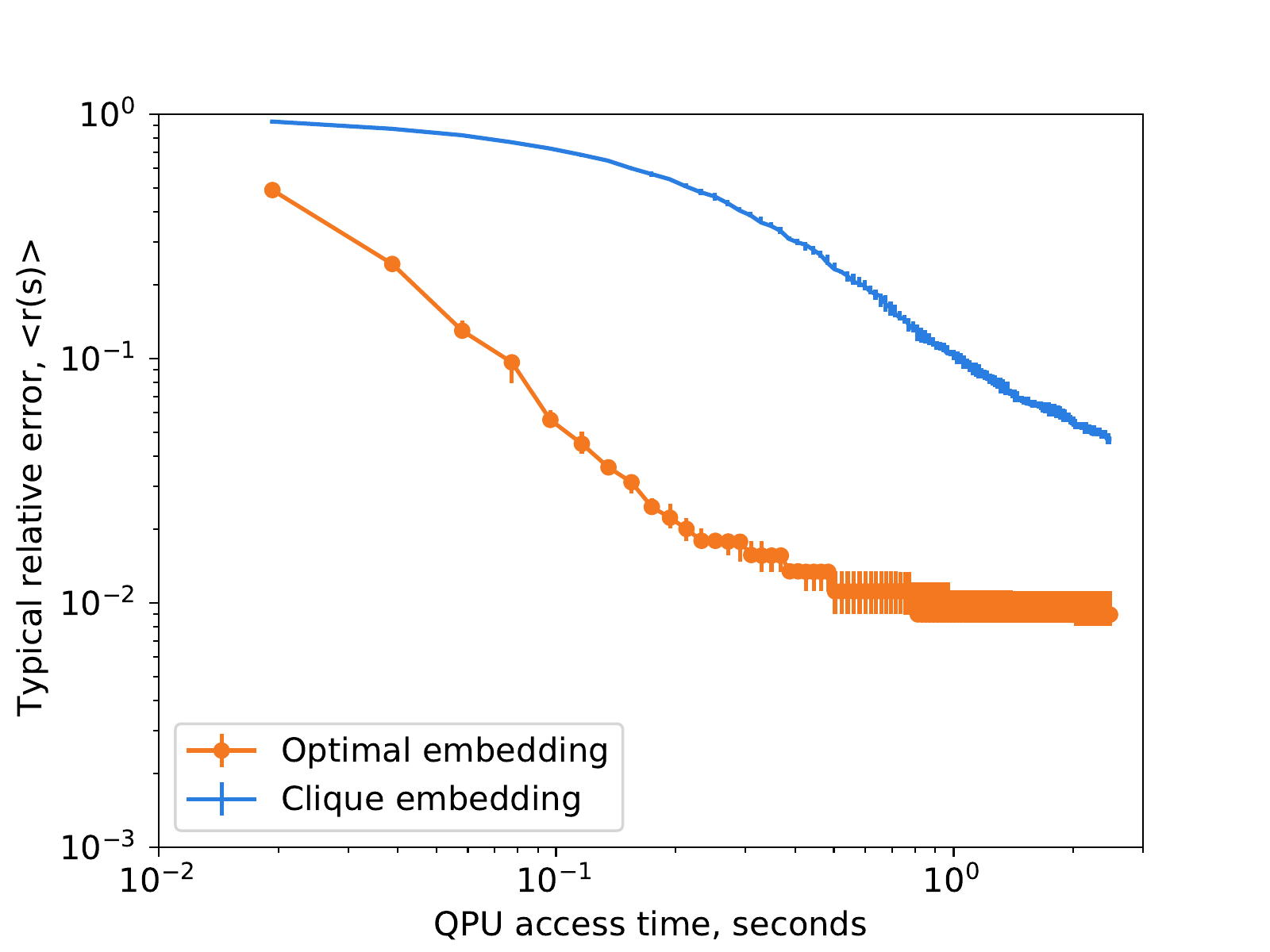}
    \caption{\label{embeff1} Hybrid method applied to a $10\times10\times10$ cubic lattice exploiting either clique or optimal embeddings, with a D-Wave 2000Q subsolver. Median error is plotted per iteration of the algorithm, with each iteration weighted by the QPU access time. Points are linearly interpolated on the logarithmic scale, error bars are standard (68\%) confidence intervals on the median. The optimal embedding method outperforms the clique embedding method on both short and long timescales.}
\end{figure}

We can first examine the consequences of inefficient embedding choices. We can analyze the method using a D-Wave 2000Q QPU as the subsolver. For this QPU a $10\times10\times10$ cubic lattice is approximately twice the programmable scale when using an optimal embedding. However, a clique minor contains at most 64 variables, sufficient only for a $4\times4\times4$ subspace. By constrast, the optimal embedding with chain length of $4$ allows iteration of $8\times8\times8$ subspaces, against which we compare.
As shown in Figure \ref{embeff1}, performance is weaker using the clique embedding.
For very short timescales a factor $\sim 8$ reflecting the relative number of variables in each embedding, approximately describes the delay in burn-down. For longer timescales, the dynamics are more easily trapped by local minima when using the smaller subspace (clique) embedding method. As shown in Table \ref{table:minorembedding}, a clique-based method on an Advantage QPU might allow subsolving of regions of size up to $5\times5\times6$ ($N=180$).\footnote{Given the pattern of unyielded qubits and couplers relevant to Advantage\_system4.1 for example, $N$ is limited to $177$ variables. We can omit three corner variables from the subcube as an accommodation.} Nevertheless, we find the gap in performance between an optimal embedding and the clique embedding remains significant.

\begin{figure}[hbt!]
    \includegraphics[width=\linewidth]{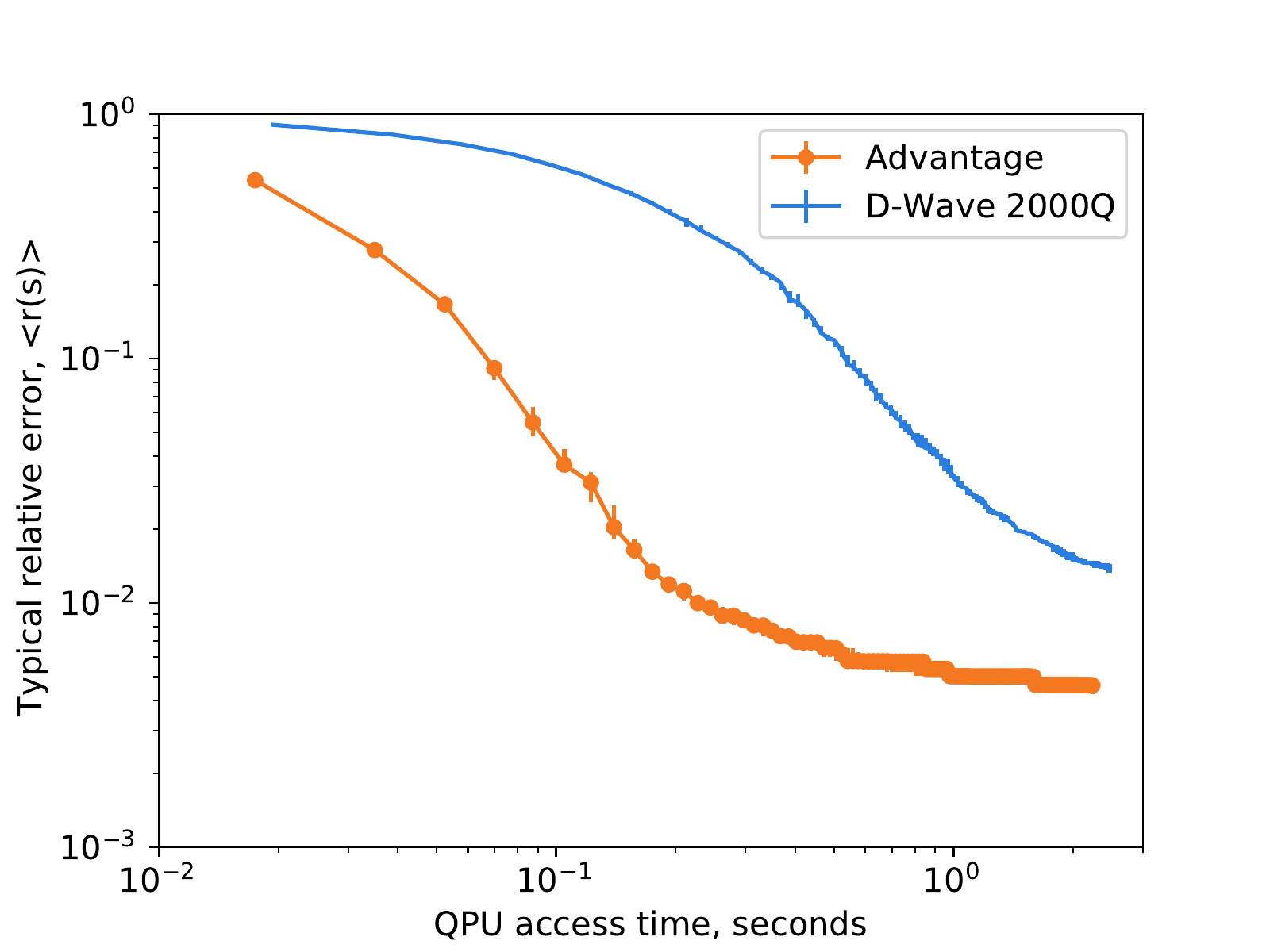}
    \caption{\label{embeff2} Hybrid method median error as a function of QPU access time, applied to an $18\times18\times18$ cubic lattice exploiting optimal embeddings for an Advantage or D-Wave 2000Q QPU. Subspaces are $15\times 15\times 12$ and $8\times 8 \times 8$ respectively for these QPUs, or rotations thereof. The Advantage QPU method outperforms the D-Wave 2000Q method on both short and long timescales.}
\end{figure}

Figure \ref{embeff2} provides a comparison of D-Wave 2000Q and an Advantage QPU-enabled hybrid methods, with optimal embedding strategies employed in each case.
The Advantage\_system4.1 solver exploits larger subspaces ($15\times15\times12$ and rotations thereof) with chain lengths of 2. This can be compared to the $8\times8\times8$ subspaces with chain length of 4 for the DW\_2000Q\_6 solver. These advantages seem most significant in explaining the performance gap, as opposed to other known QPU differences such as temperature and noise characteristics \cite{King2020}.
As discussed in Appendix \ref{app:vacancies}, subsolver yield and scale have important impact on performance --- consistent with the intuition that large area local search should only succeed in cases where subspaces have large volume and small surface area. Constraining the Advantage QPU and D-Wave 2000Q QPUs to operate as subsolvers at comparable scale and yield performance becomes similar. 

We can make a comparison to results presented by Okada et al., who also considered solutions to $10\times10\times10$ lattices on a D-Wave 2000Q solver~\cite{Okada2019}. In addition to their clique subsolver result, which closely follows ours, they considered a heuristic embedder. As shown in Appendix \ref{app:ferromagnets}, after accounting for methodological differences, the optimal embedding result of Figure \ref{embeff1} represents a significant improvement over that obtained by the heuristic embedding method. It should be emphasized that while Okada et al. present an analysis of an embedding heuristic for lattice solving, it is not restricted to such an application; it is this flexibility which is the principal cause of the performance gap.


\subsection{Comparison to alternative optimization heuristics}
A variety of classical non-hybrid methods might be employed for optimization; several are discussed in Appendices \ref{app:spin_glasses}, \ref{app:planted} and \ref{app:SA}. On short timescales, SA is among the best options~\cite{Kirkpatrick:OSA}. In this section we also examine results for a less competitive algorithm, steepest greedy descent (SGD) by single bit flips. This provides intuition for energy scales at which the landscape becomes challenging (at which trapping is possible). 

We consider SA exploiting a temperature schedule progressing geometrically from a conservative fast mixing temperature to a temperature where excitations are rare. Given the temperature bounds and schedule form, the algorithm is parameterized by a number of samples $n$ and number of sweeps $S$. 
We evaluate performance of SA using the open-source dwave-neal implementation~\cite{ocean-SDK}, which is suitable for Hamiltonians without restriction to particular lattices or coupling precision. For a set of exponentially spaced values of $nS$ we collect data, varying also the ratio $n/S$ on an exponential scale. In this way we capture a broad range of operational timescales, and can determine the optimal tradeoff between restarts and sweeps per samples. Further method details are provided in Appendix \ref{app:SA}, where it is also shown that $n=1$ is an optimal choice for minimization of the median energy on the timescales presented ($t<$\SI{100}{s}).

\begin{figure}[hbt!]
  \includegraphics[width=\linewidth]{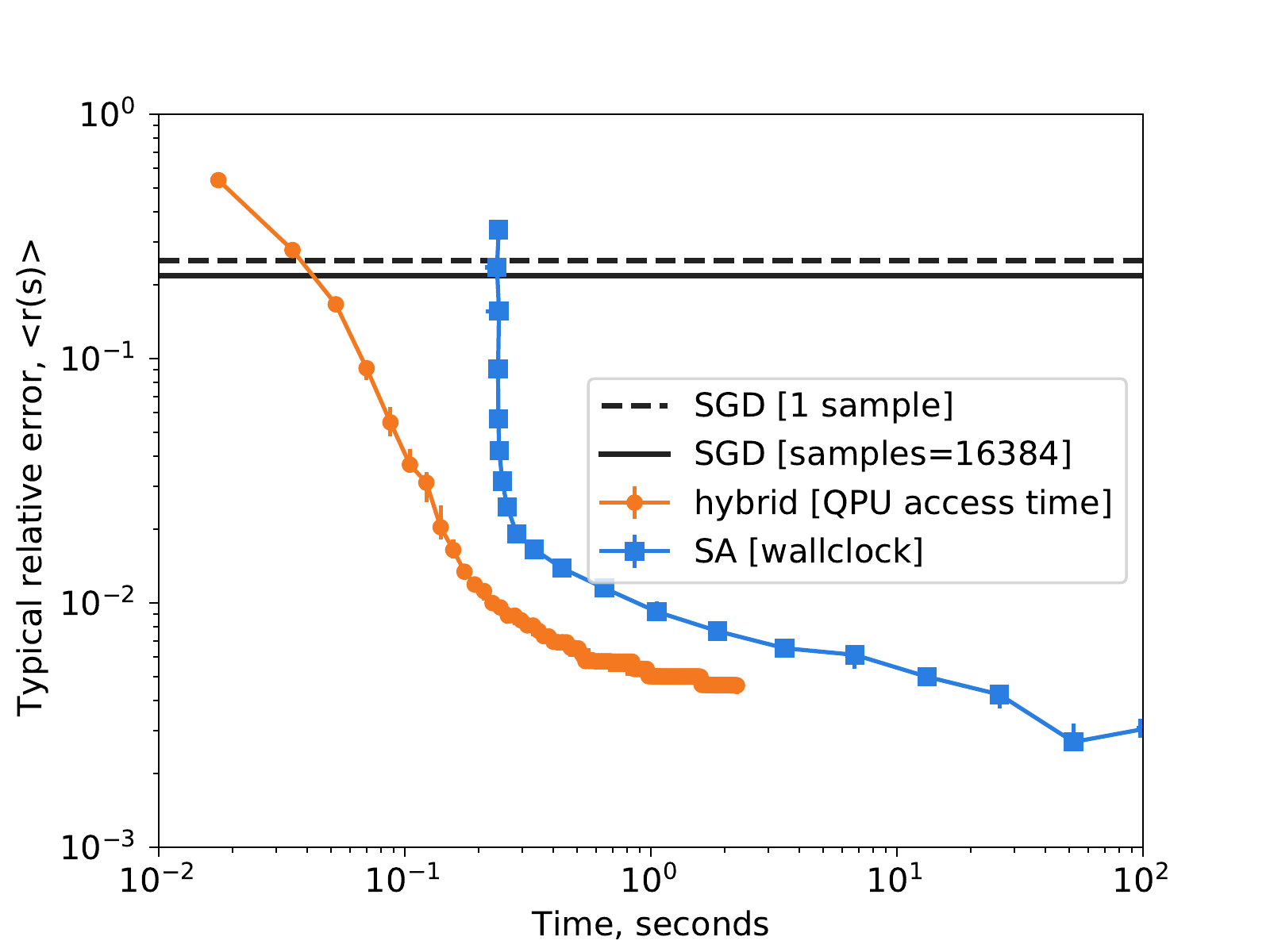}
  \includegraphics[width=\linewidth]{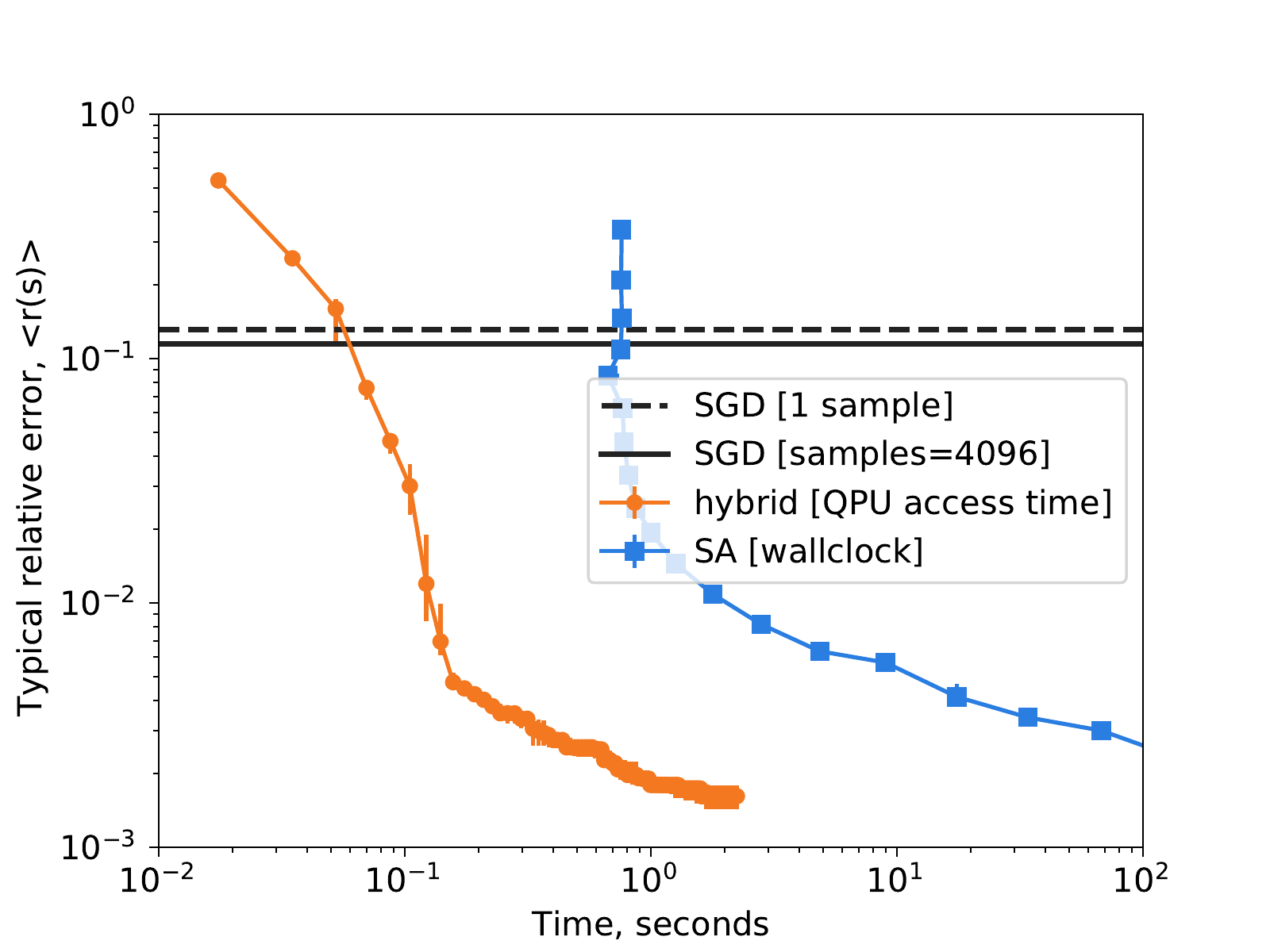}
  \caption{\label{FigVersusSA} Performance of various algorithms, energy versus time for (top) cubic periodic lattices at a scale $N=5832$ and (bottom) toric-Pegasus lattices at a scale $N=11616$. The QPU-enabled hybrid method energy is evaluated at each iteration with associated QPU access time measured. The same set of instances are evaluated by SA using $n=1$ reads, and $S=2^x$, each point representing sequential integer $x$ values beginning at $x=0$. A median is plotted in both cases with respect to instances. Comparing QPU access time to wall-clock time, the QPU-enabled hybrid method can outperform SA for a range of relative errors (equivalently, timescales). SGD is typically trapped at a large relative error by local minima. Allowing for large number of restarts, SGD cannot attain low relative errors on practical timescales. Both curves are expected to be monotonic with respect to time up to sampling error, the apparent uptick in the final SA point is a statistical fluctuation. Using no restarts ($n=1$) is optimal for the timescales demonstrated, as shown in Figure \ref{fig:convexhull}, but for longer timescales, beyond \SI{100}{s}, using $n>1$ (restarting, and keeping the best sample) becomes optimal.}
\end{figure}

For large $n$ and $S$, the number of spin updates ($n S N$) becomes approximately proportional to the run time of the algorithm and can be used to understand the efficiency of the implementation. We show in Table \ref{SpinUpdatesPerSec} the wall-clock time required for our SA implementation. For sparse lattice problems with floating-point fields and couplers, at large $S N$ and $n=1$ (regime of principal interest), dwave-neal is appropriately optimized. Specific problem structure and coupling precision restrictions might be leveraged to obtain some further gains~\cite{TroyerSTA}. Alternative hardware platforms such as GPUs can also be leveraged exploiting lattice structure~\cite{King2017,QUBITS2021}. 
\begin{table}[htb!]
  \begin{center}
    \caption{\label{SpinUpdatesPerSec} Time per spin update (wall-clock time normalized by $n S N$, with $n=1$, $S=2^{18}$) for the largest lattices considered: toric-Pegasus ($N=11616$) and periodic-cubic lattice problems ($N=5832$); a mean is presented with respect to $49$ instances. The larger update time per spin in Pegasus lattices is accounted for primarily by larger connectivity, and in part by larger scale. The implementation is single-threaded. For these operational timescales initialization is negligible. Parentheses denote the uncertainty in the final significant figures.}
    \label{tab:table1}
    \begin{tabular}{l|l|l}
      \textbf{} & \textbf{Periodic-Cubic} & \textbf{Periodic-Pegasus}\\
      \hline
      Spin update & \SI{ 33.0(3)}{\nano s} & \SI{45.5(11)}{\nano s}\\
    \end{tabular}
  \end{center}
\end{table}

We report wall-clock times for SA, and compare this to QPU Access Time for the hybrid method. The CPU is specifically an Intel\textsuperscript{TM} Xeon\textsuperscript{TM} CPU E5-2670 0 @ 2.60 GHz, and the process is single-threaded.\footnote{Intel and Xeon are trademarks of Intel Corporation.}
The hybrid method has excellent performance across a variety of timescales, as shown in Figure \ref{FigVersusSA}. Initialization time for SA is large (whereas hybrid initialization time is not included), but this has little impact on results at small relative error that are of primary interest. There is scope to improve SA by modification of the anneal schedule shape, or the compute platform, whereas the restart rate $n/S$ is suitably optimized for the timescales evaluated. The QPU parameterization for this plot is chosen to crudely account for QPU access time overheads in D-Wave 2000Q and Advantage QPUs as described in Appendix \ref{app:tsQPUp}. Results therein show that QPU subsolver performance can be improved by optimization of parameters. Variations on the hybrid workflow can also improve performance, a limited set of variations are evaluated in Appendix \ref{app:python}. 

As an additional point of comparison we show the result for SGD, specifically the open-source dwave-greedy implementation on the same CPUs used for the SA experiment~\cite{ocean-SDK}. This proves not to be a competitive method, since the algorithm is easily trapped by local minima, limiting the relative error attainable. As shown in Figure \ref{FigVersusSA}, even with large sample parallelism the method (take many samples and keep the best) is uncompetitive with annealing and hybrid methods. Obtaining this level of parallelism requires 166 seconds (16384 samples, in cubic lattice case) and 238 seconds (4096 samples, in toric-Pegasus case). A more efficient implementation might track and return only the best sample sequentially, bypassing the dimod-sampler (Python) interface that is a limiting factor in the speed of implementation. Nevertheless, it can be established that even optimized greedy methods are uncompetitive with SA and hybrid QPU methods except at high relative error.

A final classical method we might consider is SA-enabled hybrid (Algorithm \ref{pseudo}, substituting SA for the QPU subsolver). As shown in Appendix \ref{SAsubQPU}, the SA-enabled hybrid method is slower than SA applied directly to the full lattice (as shown Figure \ref{FigVersusSA}) when equating the total number of spin updates in the two methods. Furthermore, when SA is substituted for the QPU with time resources approximately matching the QPU access time, the QPU-enabled hybrid method is shown to produce lower energies. 

\section{Conclusions}

A simple greedy large neighborhood local search is proposed in which a QPU does almost all the work. It is shown to effectively sample low-energy states in spin-glass models at twice the scale directly programmable on a QPU for two different lattice classes.
We present the algorithm in open source with dwave-hybrid, sufficient to reproduce results, alongside special case proof of concept optimized code.

SA represents stiff competition in the space of spin glasses, when the task of energy minimization is performed on short timescales for large lattices.
Performance with Advantage QPUs used as the subsolver, evaluated in terms of QPU access time (accounting for programming time, readout and annealing time) shows promising performance relative to SA.
Choice of embedding, QPU yield, and QPU generation (Advantage versus D-Wave 2000Q) is shown to strongly impact performance. Tuning of the workflow and associated parameters allows scope for further improvement, as do planned developments in QPU technology.

\bibliography{hybridlattice}

\appendix

\section{Problem definitions}
\label{app:problem_definitions}
\subsection{Periodic lattices}
\label{app:periodic_lattices}

Results for simple cubic and toric-Pegasus lattices are presented in this paper. Periodic boundary conditions are chosen for ease of analysis, eliminating certain boundary effects. However, the method presented generalizes beyond these special cases to a variety of lattices for which efficient QPU embeddings are possible. 

In the periodic cubic lattice of dimension $L$ we consider a variable to be indexed as $(i_1,i_2,i_3)$ with $i_x=0\ldots L-1$, and connects modulo $L$ to $(i_1\pm 1,i_2,i_3)$, $(i_1,i_2\pm 1,i_3)$ and $(i_1,i_2,i_3\pm 1)$.
Node degree is 6 for all variables, and the number of variables is $L^3$. Among other symmetries, in such a graph a displacement in any of the three dimensions is an automorphism, node degree is 6.

The toric-Pegasus lattice of dimension $L$ we consider is closely related to the standard Pegasus graph $P[m]$ of dimension $m = L+1$~\cite{Boothby:19}. Variables are indexed as $(u,w,k,z)$. To create the periodic graph of dimension $L$ from the standard Pegasus graph $P[m=L+1]$ periodic boundary conditions $(u,w,k,z) = (u,w,k,z+L)$ and $(u,w,k,z) = (u,w+L,k,z)$ are enforced. In practice this means that we take the $P[m=L+1]$ graph, and contract $(u,L,k,z)$ to $(u,0,k,z)$,\footnote{The variable $(u,0,k,z)$ inherits the neighbors of $(u,L,k,z)$, and variable $(u,L,k,z)$ is eliminated.} and we add external couplers ($(u,w,k,0)$ to $(u,w,k,L-1)$) that span the boundary~\cite{optimized-code-placeholder}.
Node degree is 15 for all nodes and the number of variables is $24 L^2$.
Among other symmetries, a vertical or horizontal displacement of coordinates is an automorphism.

\subsection{Spin glasses}
\label{app:spin_glasses}

Spin glasses have been a staple of annealing benchmarking studies since the inception of annealing methods, and have regularly been examined in the context of D-Wave QPUs and competitor technologies~\cite{doi:10.1126/science.284.5415.779,Katzgraber2014,PhysRevE.104.025308,Kadowaki:QA,Nishimori2001,PhysRevE.92.013303,Okada2019,Harris2018,King2020,Raymond2020,Liu2015,PhysRevLett.115.077201}.
The key component in spin-glass models is a significant level of frustration, and standard ensemble definitions achieve this in a concise manner by sampling of independently and identically distributed zero-mean coupler values, at zero external field. Much is known about the phase diagram, and low energy solution space features in the large system size limit for canonical cases such as the $\pm J$ ensemble. A convenient feature of spin glasses is that they have a free energy which is self-averaging, meaning that the macroscopic properties (underlying algorithmic hardness) vary little from instance to instance. For larger systems, we can anticipate all instances to have a performance close to the median making this a good summary statistic, as used herein. 

We have considered the standard $\pm J$ coupler ensemble over lattices, in which couplers take only two values, the other most common spin-glass ensemble has normally distributed coupler values. Theory suggests that asymptotic (large $N$) properties of standard spin glasses are weakly dependent on the choice of $J_{ij}$ distribution~\cite{Nishimori2001}, including qualitative algorithmic performance across a variety of algorithm classes. The $\pm J$ ensemble is therefore a reasonable test case.

In the case of the QPU, the programmable energy scales ($J,h$) are bounded, and the choice of $\pm J$ is particularly convenient for accessing the maximum programmable energy scales~\cite{docs}, mitigating for noise. Exploiting discrete energy levels and lattice structure present in a $\pm J$ ensemble has also been demonstrated in classical software frameworks~\cite{TroyerSTA},  though these optimizations tend to be fragile to small changes in the lattice definition or coupler values, and yield only prefactor-type speed ups. Digital solvers such as dwave-neal are typically constructed to be robust to a variety of coupler distributions. By contrast, changes to our hybrid method would need to be considered for a QPU-enabled hybrid method to tolerate large variability in coupler or external field strength owing to precision limitations. 

Cubic lattice spin glasses have finite temperature, and finite transverse field, spin-glass phase transitions~\cite{doi:10.1126/science.284.5415.779,Harris2018}. Exact optimization is limited to small scales and is also slow~\cite{KolnServer}, and inference in 3D spin glasses have been of sufficient theoretical importance to justify development of special-purpose supercomputers~\cite{Baity-Jesi2017}. Annealing and other thermal and quantum evolution-based heuristics can perform well, the second order nature of the phase transition(s) is amenable to this. 

Pegasus lattices are a specific and not much studied ensemble with regards to spin glasses~\cite{Boothby:19}. Optimization in Pegasus lattices is expected to be asymptotically simpler than in 3D lattice, owing to the cellular nature of the lattice; phase transition properties might be assumed to be qualitatively similar to Chimera and some other 2D lattices~\cite{Katzgraber2014}. However, at QPU programmable scales Pegasus (and Chimera) lattices already present a formidable challenge --- given the embedding efficiency, they can be more challenging than cubic lattices at QPU programmable scale.

Some classical heuristics have demonstrated strong performance beyond annealing in lower dimensions, including 3D~\cite{PhysRevE.92.013303,PhysRevLett.115.077201}, and additional methods have been proposed for leveraging structure in QPU architectures, in particular the cellular-level structure of Chimera graphs~\cite{SelbyGITHUB,Katzgraber2014,PhysRevE.104.025308,Boothby2021}.
Many of these heuristics were applied successfully to Chimera, but most do not generalize for practical purposes to Pegasus since they involve operations that scale exponentially in the local connectivity, which is much higher and less regular in Pegasus~\cite{Boothby:19, Zephyr}. Other methods typically involve the use of multiple replica (iteration of more than one samples in parallel). This replica overhead can, in our evaluation, make the methods inferior to SA on the shorter timescales presented even if they offer superior performance on longer timescales and in harder instances.\footnote{Some methods such as population annealing might include SA as a limiting case of the parameterization. Requirements for ensemble-wise, or instance-wise, tuning of additional parameters is another reason we exclude more complicated methods from our analysis.}

The performance of QPUs for cubic lattices at programmable scales has been evaluated as a function of annealing time~\cite{Harris2018,King2020}. Many programmings are necessary to reliably optimize problems at maximal programmable scale, which is informative on the limitations of a QPU as a subsolver. Pegasus spin glasses (with standard non-periodic boundary conditions) are also challenging to solve to optimality with QPUs at maximum programmable scale. Subproblems studied in the hybrid case are slightly different in that they are aided (or hindered) by conditioning information: random external fields in spin glasses typically make problems easier to solve, and as the ground state is approached these external fields may become correlated promoting a clear ground state. Furthermore, the task we present is different and easier than obtaining the ground state with high confidence: merely to obtain the lowest energy possible on some constrained timescale. One programming returning few samples per iteration may suffice, since attainment of a suboptima can allow for progress of the energy estimate, particularly early in the algorithm.

\subsection{Ferromagnetic and planted instance results}
\label{app:planted}
Evaluation of ferromagnetic and planted-solution ensembles allows a demonstration of the algorithm attaining optimality. The cases evaluated are not as hard as spin glasses from the $\pm J$ ensemble, but allow some greater transparency of behaviour at low energies, as well as providing a contrast to spin-glass results.

\subsubsection{Ferromagnets: Comparison to \cite{Okada2019}}
\label{app:ferromagnets}
For comparison against previous studies, and for intuitions sake, a useful example is a ferromagnetic model. In this case the target Ising model is not computationally challenging because it is unfrustrated, with $J_{ij}=-1$ (ferromagnetic) for all lattice edges. The ground state has all spins aligned, and in the cubic case the ground-state energy per spin is precisely $-3$. 

Despite the fact that ferromagnets are unfrustrated problems, greedy descent from random initial conditions cannot efficiently solve large lattice problems owing to phase coexistence. Similarly, LNLS can become trapped by local minima in principle, and in practice this would be expected where subsolvers are much smaller than the full problem scale. For the case of ferromagnets it is algorthmically straightforward to work around this issue, by use of cluster algorithms or nucleation strategies, for example~\cite{Rams2016}. Nevertheless, it should be a good model for certain types of non-trivial algorithmic barriers to optimization. 

\begin{figure}[htb!]
  \begin{center}
    \includegraphics[width=\linewidth]{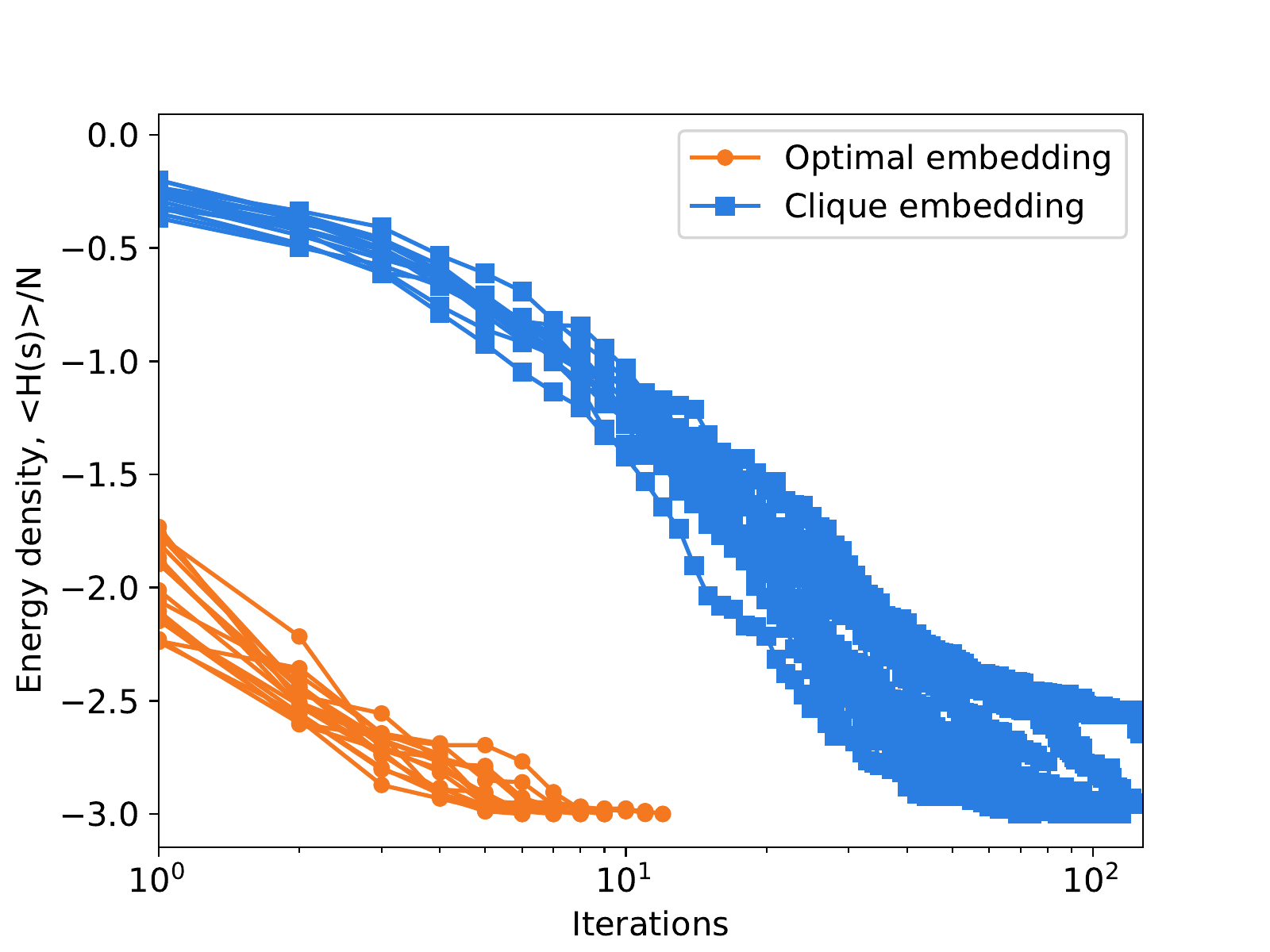}
    \caption{\label{fig:ferro} Eight independent runs (uniform random initial conditions, subspace and sample-set sequences) of the hybrid QA algorithm are shown for a $10\times10\times10$ ferromagnet, using either a fully connected embedding of $4\times4\times4$ sublattices, or an optimal embedding of $8\times8\times8$ sublattices.  The latter minor embeddings allow faster convergence to optimality, and with favorable performance against efficient embedding heuristics requiring approximately 30 iterations to optimality~\cite{Okada2019}. Energy density is related to residual error in this case as $\langle H(s)\rangle/N = 3(\langle r(s)\rangle-1)$: the ground state energy is $-3 N$.}
  \end{center}
\end{figure}

In Figure \ref{fig:ferro} we show convergence to the ground state for our method using a D-Wave 2000Q QPU in a format compatible with Okada et al.\ Figure 3~\cite{Okada2019}. Our default methodology involves no post-processing,\footnote{Majority vote applies in the cubic case, as the solver default. For Pegasus lattices this is irrelevant with chain length of one. For cubic lattices embedded on Pegasus we use the default majority vote post-processing, but since chain length is two, this is for practical purposes the same as random assignment for invalid chain states.} by contrast the method of Okada et al. involves a greedy post-processing step. Other differences exist in the specific solver used and parameterization thereof; our non-defaulted parameters are described in Appendix \ref{app:parameterization}.

Figure \ref{fig:ferro} shows that the algorithm employing the optimal embeddings quickly converges to optimality in approximately 10 iterations, whereas the algorithm employing the fully connected methodology converges to the optima on a timescale of 100 iterations. This clique solver timescale to near-optimality is close to the result presented by Okada et al., and curves are made closer with the addition of a greedy descent post-processing component.
The heuristic minor-embedding algorithm of Okada et al.\ also achieves optimality, but on a slightly longer timescale of approximately 30 iterations, significantly slower than with the optimal embedding, in spite of the additional post-processing workflow element. A similar qualitative gap can be established for the spin-glass case presented in Figure \ref{embeff1}. We conclude that performance is improved in using optimal embeddings relative to either the graph-agnostic heuristic proposed by Okada-san et al., or the clique embedding.

\subsubsection{Planted instances: Attainment of optimality in non-trivial models}

To create a problem class with a known ground state, and non-trivial patterns of frustration we can use the tile planting scheme proposed by Hamze et al. for cubic lattices~\cite{Hamze2017,Perera2020}. In the ensemble we analyze, tiles are chosen randomly subject to the constraint that in expectation the mean ground-state energy is $-1.8 N$. In this way we crudely approximate the level of frustration characterizing the $\pm J$ spin-glass ensemble, where the ground state energy is also close to $-1.8 N$.
We call this the tile-planted ensemble.

Specifically we note there are 63 different $2\times 2\times2$ cubic tiles that might be chosen consistent with some planted ground state. Each tile $i$ is associated to a ground-state energy contribution $-12 \leq e_i \leq -6$, depending on the number and configuration of frustrating couplings. We sample tiles from a distribution $P(i)$ that maximizes entropy $-\sum_{i} P(i)\log P(i)$, subject to the constraint on the ground state energy $-1.8 = \sum_i e_i P(i)/12$. About $44\%$ of tiles are maximally frustrated (3 frustrated bonds, so called F6) in a typical instance drawn from this ensemble, with a tiny fraction also of unfrustrated cubes, and a variety of tile types in between.
The problems generated have couplers $\pm 1$ as in the spin glass, the ground-state energy is approximately equal to a spin glass, and the distribution of coupler values is locally indistinguishable from a spin glass.~\footnote{By evaluating only the coupling distribution, or distribution over subtrees, one cannot distinguish a standard $\pm J$ spin-glass instance from a tile-planted instance.}

\begin{figure}[htb!]
  \begin{center}
    \includegraphics[width=\linewidth]{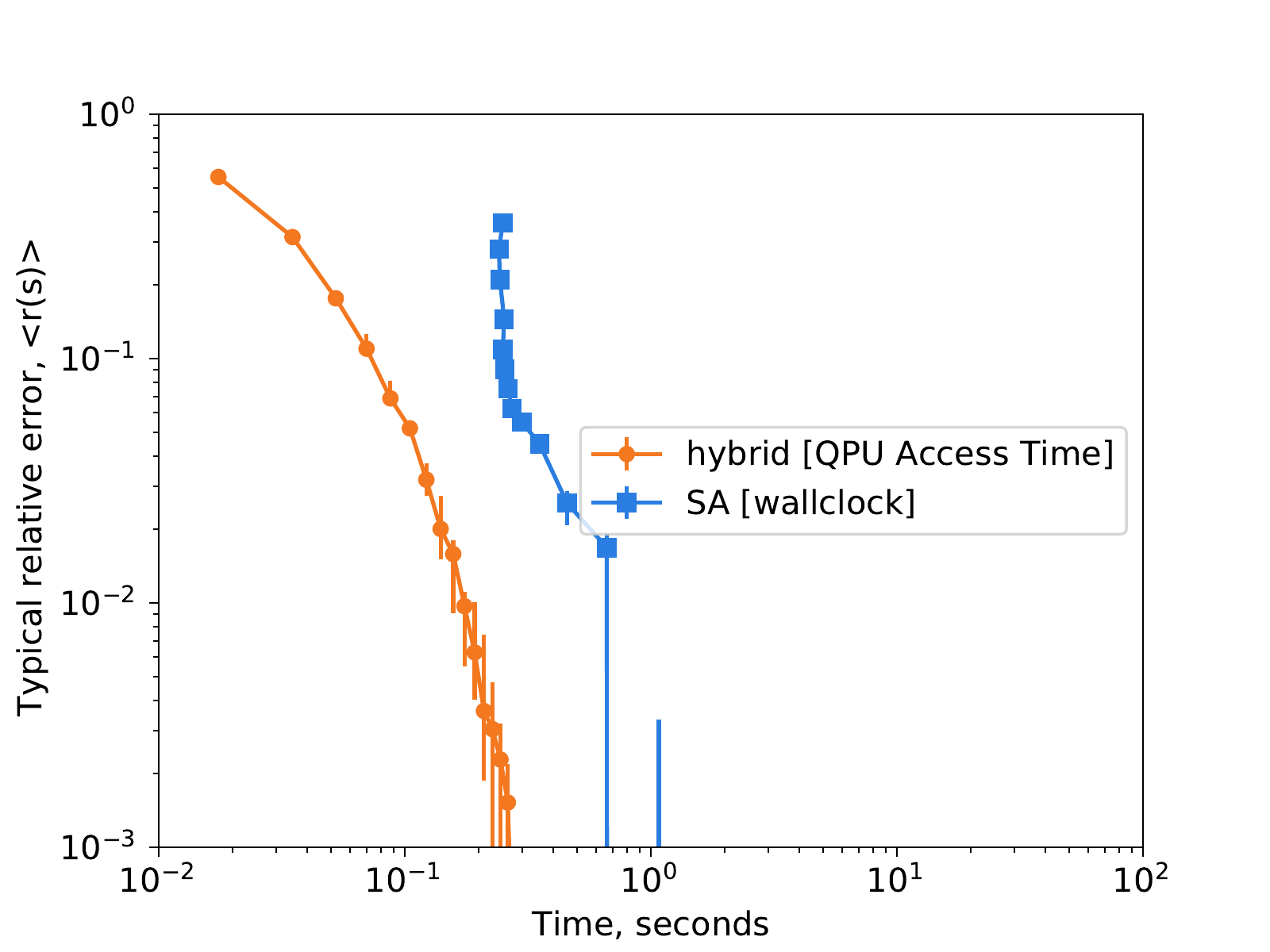}
    \caption{\label{CTile1-F1p8} Median relative error achieved over $18\times 18 \times 18$ periodic-cubic lattices. The ground state $E_0$ is known by construction, and in the median is achieved by both SA and the QPU-enabled hybrid algorithms on short timescales. The residual error curves collapse to zero on these timescales -- within \SI{0.3}{s} of QPU access time for the hybrid method, and at approximately \SI{1}{s} for annealing implemented with dwave-neal (using $n=1$). }
  \end{center}
\end{figure}

These problems are significantly more challenging than a ferromagnet, but we find that for all heuristics and exact solvers they are in typical case much simpler to solve than $\pm J$ ensemble spin glasses. Our standard hybrid workflow quickly and reliable achieve an optimum for planted problems, as shown in Figure \ref{CTile1-F1p8}. It should also be noted that simulated annealing achieves optimality, on a timescale much shorter than for the $\pm J$ spin glass, although again longer than for ferromagnetic cases. Despite the strong performance of the QPU-enabled hybrid method and SA, section \ref{sec:complete} indicates that the pattern of frustration is not easily amenable to all optimization methods, thus we regard this as an interesting test class. We have shown results for a specific distribution of tiles. The cubic lattice planting method allows for creation of more frustrated (higher expected energy) ensembles, which are harder to solve~\cite{Hamze2017}. We find that these methods typically fail to achieve optimality on comparable timescales in highly-frustrated cases. 

\subsubsection{Performance of complete solvers}
\label{sec:complete}
It is useful to understand the capabilities of complete solvers, those which provide a certification of the ground state, for the problem classes we evaluate.
Complete solvers are a common choice for many optimization tasks, and use distinct methodologies from annealing, such as mixed integer programming.
Gurobi\textsuperscript{TM} run in a default mode is chosen as an exemplar for this class.\footnote{Gurobi is a trademark of Gurobi Optimization, LLC.}
For spin glasses matching QPU architectures, such as Chimera, complete solvers have been shown to be slow~\cite{McGeoch2013,Coffrin2017}. Complete solvers might be tuned to special lattice cases such as cubic lattices~\cite{KolnServer}, but operation in these cases is also restricted to scales much smaller than those examined in this paper. It is expected that complete solvers will be slow for a variety of canonical spin glasses, in particular periodic-cubic or toric-Pegasus $\pm J$ ensemble instances.  Many planted and weakly frustrated ensembles are not challenging for complete solvers~\cite{Coffrin2017}, ferromagnetic structure for example allows for efficient inference and verification of a ground state. 

\begin{figure}[htb!]
  \begin{center}
    \includegraphics[width=\linewidth]{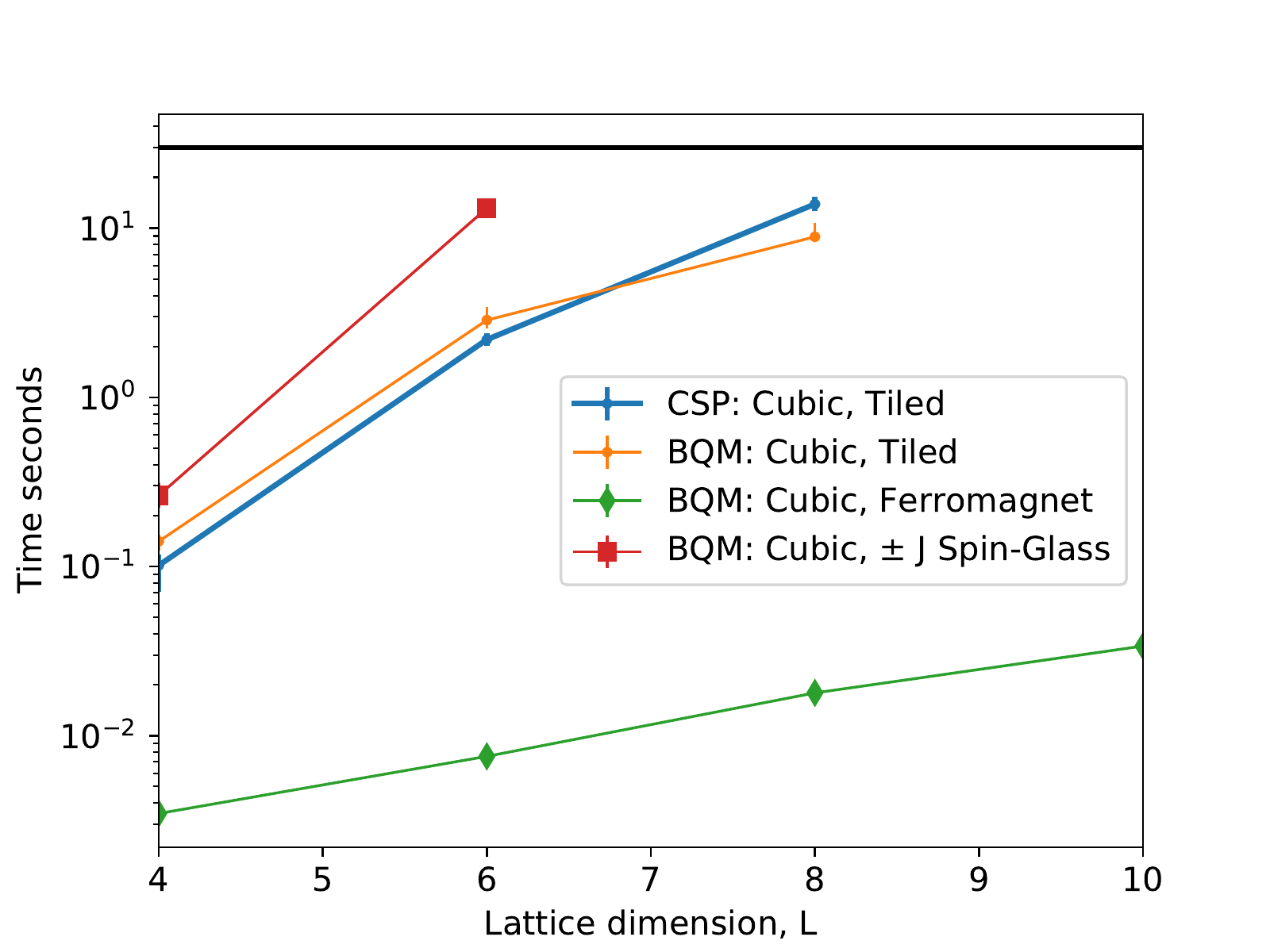}
    \caption{\label{Gurobi} Median Gurobi solver time to solution for the tile-planted, $\pm J$ spin-glass, and ferromagnetic ensembles on periodic $L\times L\times L$ cubic lattices. A time-out of 30 seconds is applied for each instance; medians exceeding this (marked) threshold are excluded from the plot. At scales studied, complete solvers like Gurobi are not competitive under default operation for tile-planted and $\pm J$ spin-glass instances. By contrast ferromagnets are solved by Gurobi with ease. }
  \end{center}
\end{figure}

The optimization task for tile-planted instances can be formulated either as an Ising model (per usage in hybrid and annealing heuristics), or as a constraint satisfaction problem. Each tile represents a constraint over 8 variables. When a problem is presented in the form of a set of constraints, we reveal the location and type of every tile, extra information that one might assume makes problem-solving easier. We can consider a sequence of $L \times L \times L$ problems varying $L$, and determine the time for Gurobi to return a proof of optimality under default operation. Gurobi operates via Amazon Web Services, and we report the solver time, results are shown in Figure \ref{Gurobi}. Gurobi very quickly solves ferromagnets up to large sizes, but by contrast, with a time-out of 30 seconds cannot solve spin glasses at a scale $L=6$, or tile-planted instances at a scale $L=8$. Gurobi is impractical for problem scales of $L=18$ on which we focus. 
Interestingly, presenting the problem in the form a constraint satisfaction problem does not appear to significantly accelerate solving, an indication that the planting structure does not trivially imply the solution.
This is consistent with literature results on these tile-planted instances, which are hard for a variety of heuristics when frustration is large~\cite{Hamze2017,Perera2020}. The strong strong performance of the hybrid method and SA in Figure \ref{CTile1-F1p8} indicates that in this problem class heuristics are superior for optimization purposes.

\section{Simulated annealing}
\label{app:SA}

A given run of simulated thermal annealing (SA) is parameterized by a number of samples ($n$) and a number of sweeps per sample ($S$), and a schedule (progression pattern for the temperature, from high to low)~\cite{Kirkpatrick:OSA}. 
We consider the open-source CPU implementation dwave-neal, with a geometric schedule --- terminated by one quench sweep~\cite{ocean-SDK}.
The initial samples for the process are generated uniformly at random.
The sequence of temperatures at which Metropolis sweeps are performed is determined by a geometric schedule:\footnote{A sweep consists of the N variables being updated in a fixed order, with single-variable flips accepted according to the Metropolis Hastings Rule.}
\begin{equation}
  T = T_{max} (T_{min}/T_{max})^{s/S}
\end{equation}
for $s=1$ to $S-1$, terminating with one additional sweep at $T=0$. The quench sweep, at the end, is useful to enhance performance on short timescales. The choices $T_{max}=k/log(2)$ and $T_{min}=2/\log(100 N)$ represent conservative bounds, with $k$ the lattice connectivity:\footnote{For the case of $T_{min}$, $2$ is the minimum energy gap (since $J_{ij}=\pm 1$), $\log(N)$ is the entropic term (excitations must be suppressed everywhere), and $100$ is a somewhat arbitrary factor ensuring excitations are very rare in returned samples.}
sufficient respectively for fast mixing and low excitations rates in the final iteration. It should be understood that bounds may be tightened, and that a geometric schedule might be amended, for some performance gains. This could be informed by a class-wise parameter tuning exercise over a test set (for example) or based on phase diagram insight, but is not considered in this paper.

\begin{figure}[htb!]
  \begin{center}
    \includegraphics[width=\linewidth]{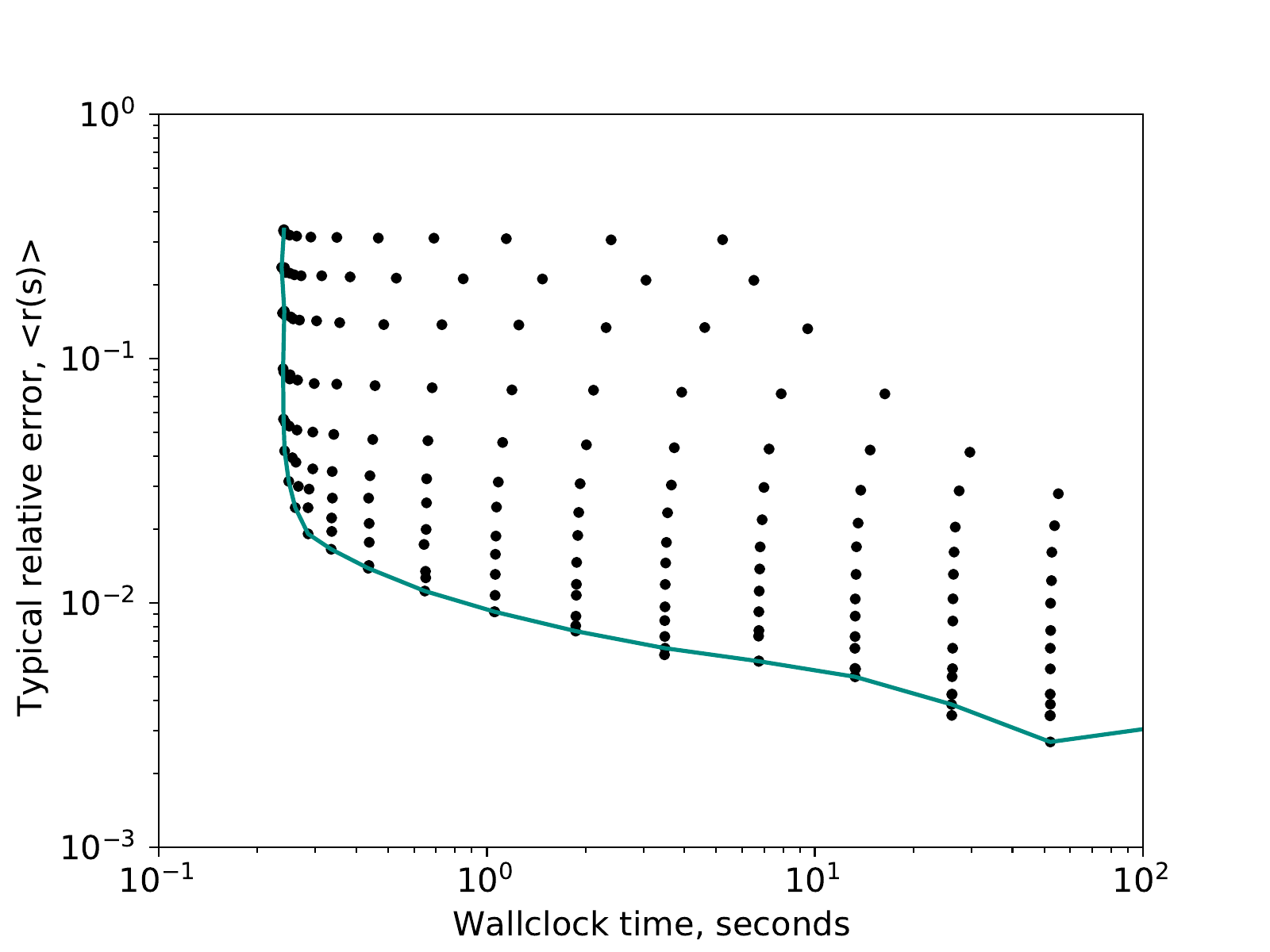}
    \includegraphics[width=\linewidth]{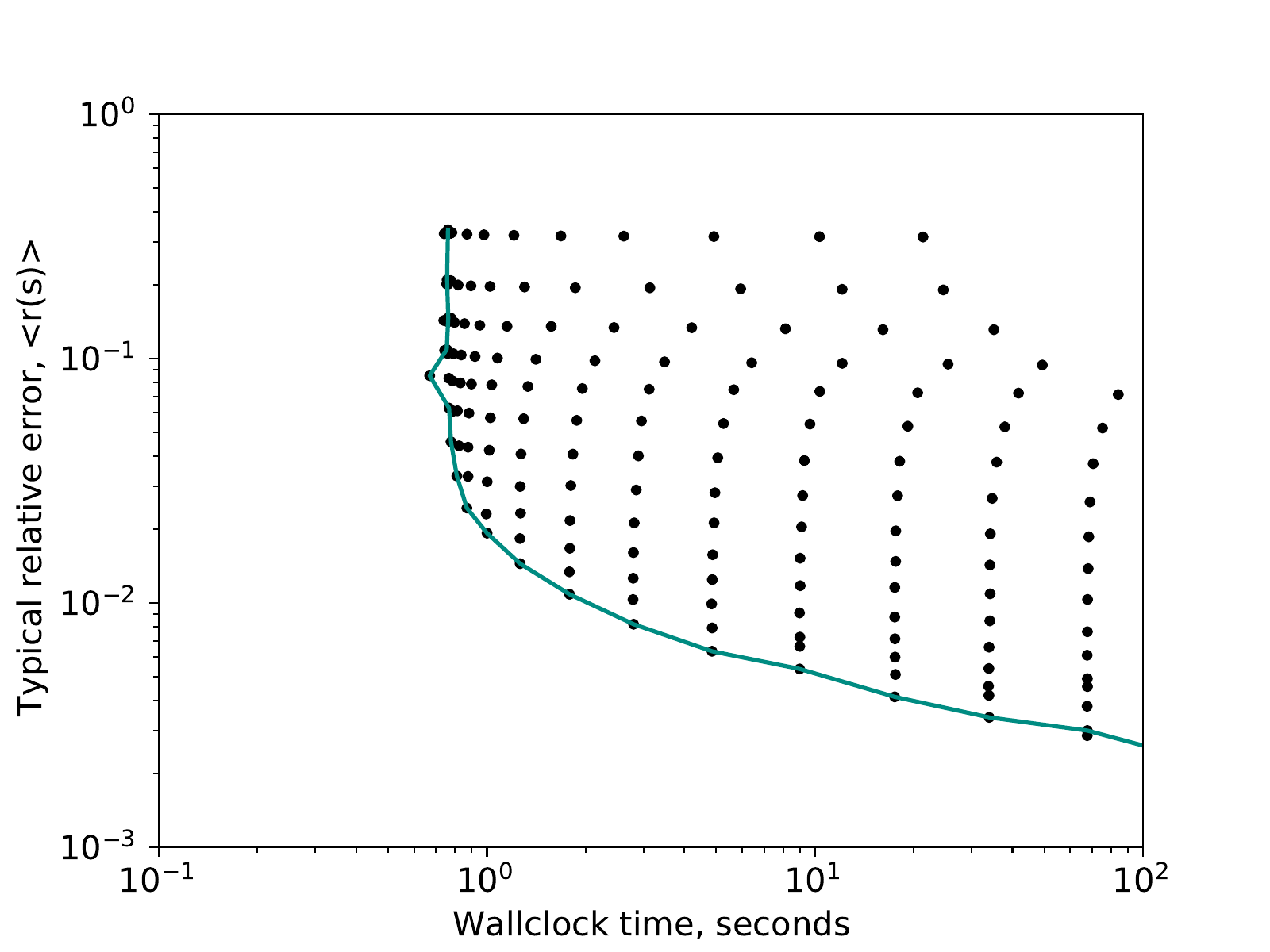}
    \caption{\label{fig:convexhull} Simulated annealing can be applied to periodic-cubic (top) and toric-Pegasus spin glasses. Median relative error with respect to $25$ instances is presented for $n=2^x$ and $S=2^y$ for integer $x$ and $y$, subject to the bound $0 \leq x+y\leq 18$ and $0 \leq x \leq 13$, where error bars are omitted for clarity. Within statistical uncertainty, the parameterization $n=1$ ($x=0$) minimizes the residual error in expectation on timescales up to \SI{100}{s} in both lattice types, though larger $n$ values were more performant than $n=1$ on some longer timescales. The more interesting part of the curve is at lower residual energy where run time is approximately proportional to $nS$, for shorter timescales initialization dominates, but this initialization is small compared to the timescales required to reach low relative error. For larger values of $n$ some inefficiencies exist in the wrapper, but this impacts only the non-performant larger relative error data.}
    \end{center}
\end{figure}

As discussed in the main text, we can operate for a variety of values of $n S$, producing energy versus time data for a broad range of parameterizations. The hull of the data is characteristic of a good parameterization tuned for all possible timescales. As shown in Figure \ref{fig:convexhull}, for timescales of up to 100 seconds pursuing a single sample ($n=1$) for largest possible $S$ typically achieves the best possible energy (describes the hull well, up to statistical fluctuations), rather than dividing compute resources between multiple samples ($n>1$). The curves presented in Figure \ref{FigVersusSA} correspond to $n=1$.

For purposes of evaluating the relative error (\ref{relativeerror}) for the $\pm J$ ensemble in all plots, we used supplementary SA runs with $n \times S$ at least two orders of magnitude larger than presented in Figure \ref{fig:convexhull}, and $S$ more than an order of magnitude longer than presented therein, though with more carefully chosen combinations of $n$ and $S$. Identical parameterizations were used for all instances to avoid introduction of a bias. From this supplementary data we determined ground-state energy estimates $E_0$ that were not improved upon in execution of hybrid methods.

\subsection{Simulated annealing as a subsolver}
\label{SAsubQPU}
The QPU-enabled hybrid method is proposed as a solution to the problem of limited QPU scale. SA implemented on CPUs does not suffer from any such severe size limitation, and can be applied in conventional form to the largest lattices we examine without hybridization, as we have done in Figure \ref{FigVersusSA}. Nevertheless, it is interesting to examine SA when employed as a subsolver, specifically to understand the performance of the hybrid method when an alternative appropriately resourced classical solver is substituted for the QPU.

To make a comparison to the QPU, let's assume the geometric schedule already described, and substitute for the QPU: using an identical subproblem scale and structure. 
Treating the startup overhead of SA as negligible, the time-per-spin update in an efficient implementation is well approximated by the values in Table \ref{SpinUpdatesPerSec}. We can then determine the number of spin-updates that might be completed by SA subsolvers in the $\SI{17.5}{\milli s}$ of QPU access time employed per iteration when using the QPU. For the toric-Pegasus lattice subsolver, working on $N=5373$ variable problems, we approximate completion of $S\sim 72$ sweeps. For the periodic-cubic lattice subsolver, which uses $N=2684$ variables, the corresponding value is $S\sim 198$ sweeps. 
The optimal way to use this many sweeps to minimize energy in SA is to process a single sample ($n=1$, no restarts).

\begin{figure}[hbt!]
  \includegraphics[width=\linewidth]{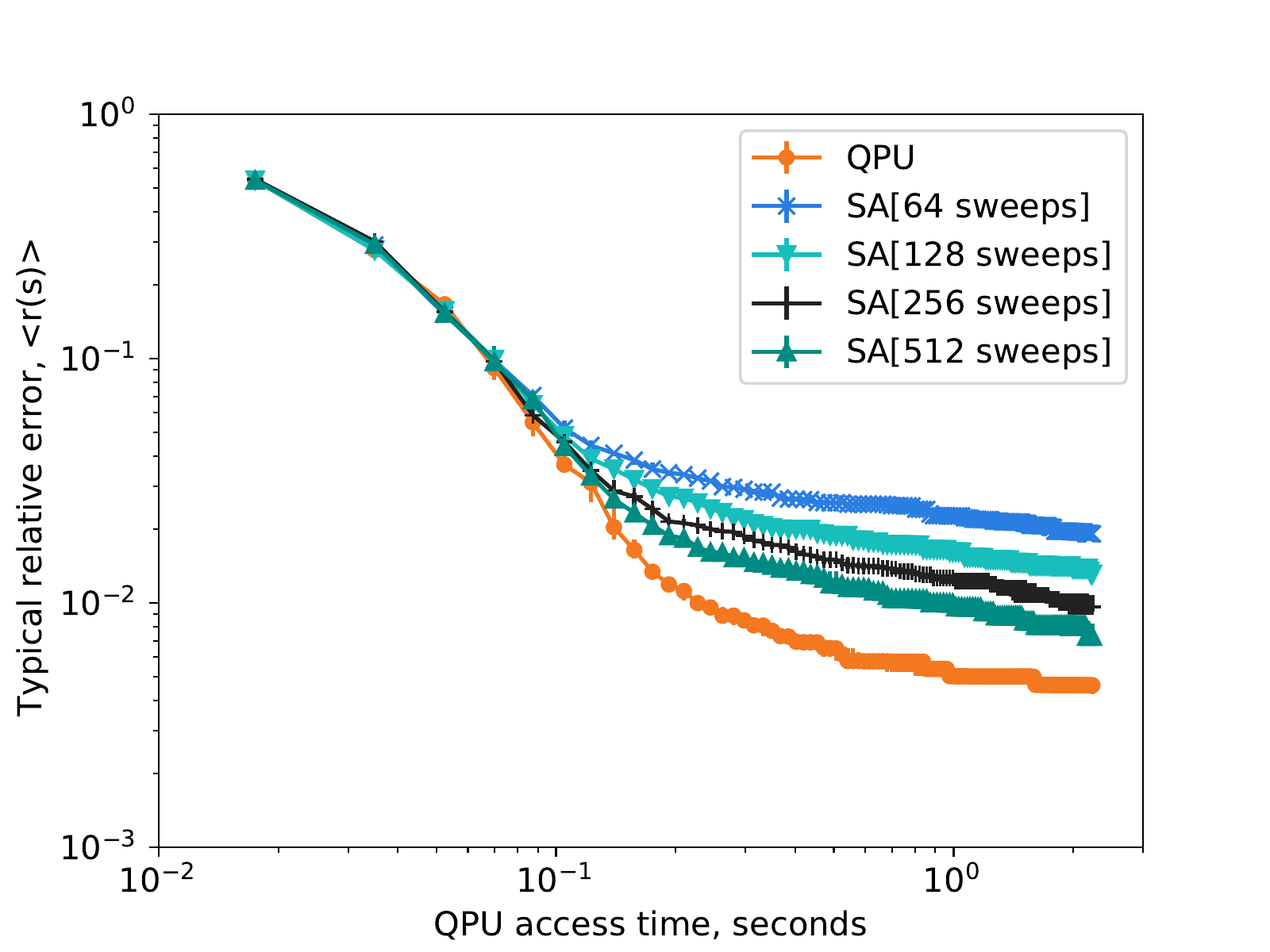}
  \includegraphics[width=\linewidth]{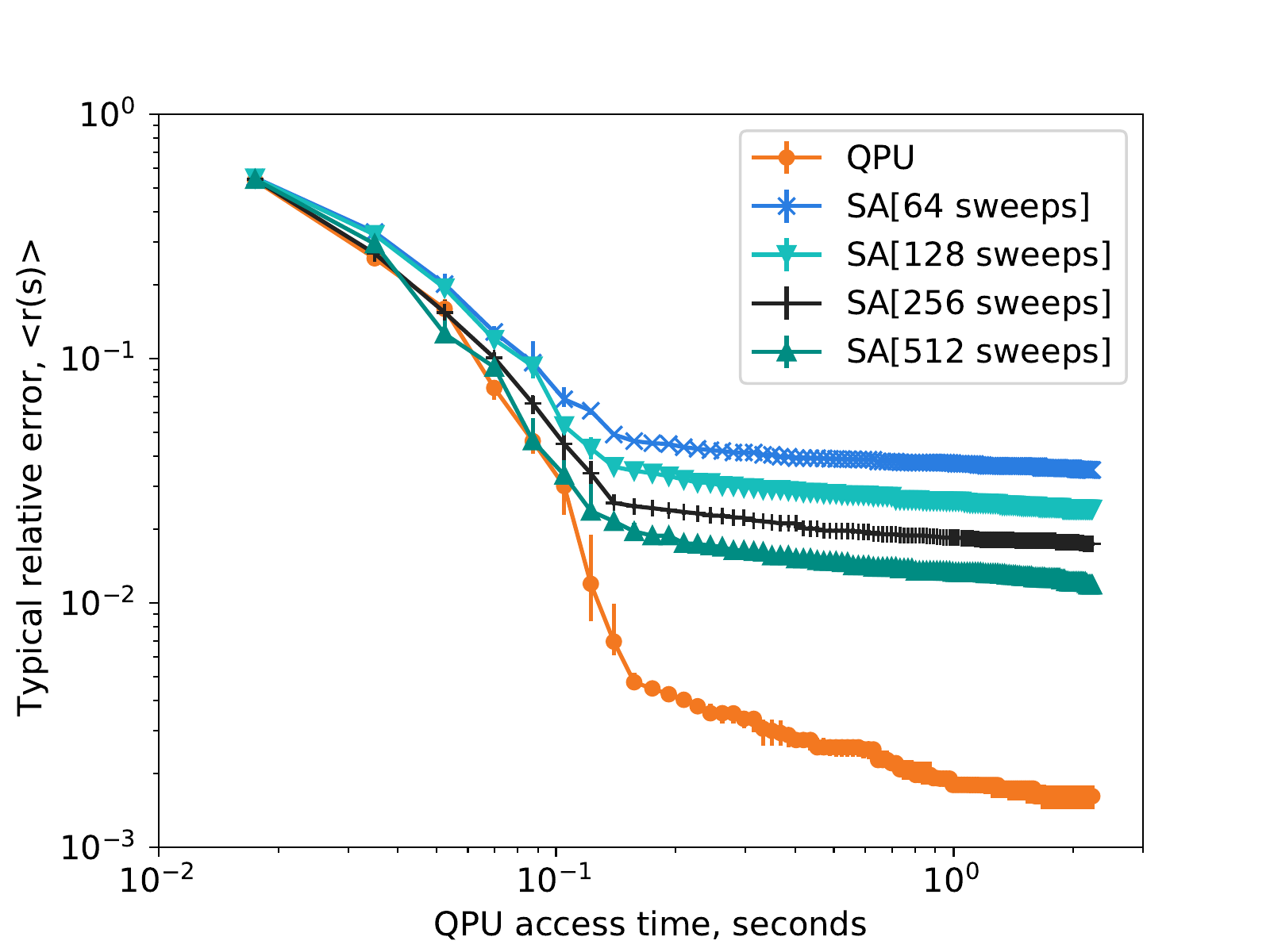}
  \caption{\label{FigVersusSAsub} Median relative error when substituting an SA subsolver for the QPU subsolver, operating with various number of sweeps. In application to the periodic-cubic lattice appication (top), ignoring initialization, an SA subsolver might complete approximately 198 sweeps per iteration based on the measured spin-update rate for large lattices. In application to the toric-Pegasus lattice the subsolver might complete approximately $72$ sweeps. At these comparison points the QPU-enabled method achieves lower residual energy}
\end{figure}

Figure \ref{FigVersusSAsub} shows the performance of the hybrid method with either the QPU as a subsolver, or with SA substituted and employing: $S = 2^x$ for $x=8,\ldots, 11$ sweeps, using a total of $2^{15}\ldots 2^{18}$ sweeps across the 128 iterations presented. Across this range of resources, the QPU-enabled hybrid method achieves lower energy.

SA also achieves lower relative error than the SA-enabled hybrid method when comparing an equal total number of spin-updates. Obtaining a relative error of $10^{-2}$ is shown in Figure \ref{fig:convexhull} to require fewer than $2^{12}$ sweeps with standard SA --- as shown in Figure \ref{FigVersusSAsub}, the SA-enabled hybrid method requires significantly more sweeps to obtain this threshold (and time, accounting for the the factor $\sim 2$ speedup of the subspace sweeps compared to the full-lattice sweeps).

\section{Hybrid implementation details}
\label{app:hybriddetails}
\subsection{D-Wave hybrid}
\label{app:dwavehybrid}
D-Wave hybrid (dwave-hybrid) is a general, minimal Python framework for building hybrid asynchronous decomposition samplers for quadratic unconstrained binary optimization problems~\cite{ocean-SDK}. In this appendix we summarise the main components relevant to our study and provide implementation examples and results supplementary to the main text presentation.

We have implemented in dwave-hybrid a reference class LatticeLNLS, to complement existing reference worflows such as qbsolv and Kerberos, supported by several new runnables (executable workflow elements) and functions. The lattice workflows require some specific additional structure relative to other reference workflows. We require as input a binary quadratic model (BQM, our Ising model being a special case) with variables specified by geometric coordinates (e.g. $(i_1,i_2,i_3)$ for the case of a cubic lattice), and at a known lattice scale (e.g. $L\times L\times L$). This is combined with a specification of the origin embedding(s), a set of good embeddings for subspaces rooted at the origin. With respect to the cubic lattice for example, a subspace is a cuboid with each of the three dimensions $\ll L$, rooted in the origin $(0,0,0)$, combined with a mapping of variables to chains compatible with the QPU-architecture.
Finally, a method to extract lattice subproblems randomly displaced from the origin is required. If the full problem is an $L\times L \times L$ cubic lattice, we can create a subproblem by choosing the origin-embedding uniformly at random among those available, and then choosing one of $L^3$ different displacements from the origin. The subspaces wrap around the boundary.

The reference example LatticeLNLS is supported by two components: the function make\_origin\_embeddings, which creates several optimal sub-problem embeddings; and the class SublatticeDecomposer, a runnable workflow component that extracts subspaces offset randomly from the origin. An origin embedding allows for some vacancies on the interior of the region, accommodating an incompletely yielded QPU, as examined in Appendix \ref{app:vacancies}.

Algorithm \ref{pseudo}, as well as the variations discussed in Appendix \ref{app:python} can be implemented in dwave-hybrid following the lattice\_lnls.py reference example~\cite{ocean-SDK}. In order to output energy statistics, QPU access time per iteration, and other runtime information, the Log runnable can be inserted into the workflows. Logging has a small impact on wall-clock clock time, since full state information is copied forward in runnables by default, but QPU access time is not impacted.

\subsection{Timing of Algorithm \ref{pseudo} excluding the QPU}
\label{app:timing}
The distributed nature of Algorithm \ref{pseudo}, employing the QPU via an API makes it challenging to understand timing, as discussed in Section \ref{results}, this relates primarily to difficulties in evaluating the QPU-API call. In addition certain elements of the dwave-hybrid package that are well adapted to prototyping and analysis of flexible workflows, are found not to effectively leverage the simplicity of Algorithm \ref{pseudo}.

A number of mappings exist in Algorithm 1. An extraction of a subproblem from the full problem, a mapping of the subproblem to programmable coupler and external field values, a mapping from the read qubit states to variable assignments, and a merging of subsamples back to full sample assignments. These mappings are informed by the optimal embedding. In addition there is the creation of the samplesets, via the QPU-API call, the evaluation of conditioning fields $h_i$, and evaluation of subproblem energies per sample.

Creation of the mappings is a one time cost per combination of lattice-type (say cubic) and subsolver (say Advantage: Pegasus $P[m=16]$ structured, with some solver-specific yield pattern). These mappings might then be reused on each iteration for a given problem, or for multiple lattice-structured problems.

Other computations must be performed on a per problem and per iteration basis. Besides generation of the samples, the most expensive floating point operations are the sparse matrix multiplications between the samples, and matrix ($J$), required for the energy calculations. Given that we pursue a methodology with few samples and modest connectivity lattices with O(1000) variables, this might be very fast compared to the QPU-API call, and associated QPU access time. 

To understand the latency more concretely we analyze a C++ implementation of Algorithm \ref{pseudo} for the cubic lattice case with an Advantage (P16) scale subsolver~\cite{optimized-code-placeholder}, but replacing the QPU-API call with a stand in function. This function is assumed to take programmable values of $J$ and $h$ as arrays in a fixed order, and returns an $n_R \times N_q$ array of random integer values $\pm 1$.
This matches close enough (for demonstration purposes) the QPU-API interface format for programming~\cite{docs}.

For both clarity and efficiency we simplify the uniform-spreading method of embedding, and the majority-vote post-processing, without significant impact on the algorithm~\cite{Raymond2020,docs}. Deviations from the dwave-hybrid reference implementation are summarised as:
\begin{itemize}
\item[1] A default approach to embedding involves, for a chain of length $C_L$ modeling variable $i$, programming the external field on each qubit as $h_i/C_L$. It is sufficient and simpler to program the first qubit in the chain as $h_i$. 
\item[2] A default approach to mapping read qubit states to variable assignments is majority vote over chains. A simpler approach is to assign the value equal to the first qubit state in the chain. Since the QPU is parameterized to have few chain breaks, and chain length is two, this method is no less powerful than majority vote for practical purposes.
\item[3] Two programmable couplers exist between the chains $(x,y,z)$ and $(x,y,z+1)$ in the optimal embedding, and it is common practice to program each coupler as $J_{(x,y,z),(x,y,z+1)}/2$. In our approach we instead program only the first of the two couplers as $J_{(x,y,z),(x,y,z+1)}$, which is sufficient.
\end{itemize}
These changes allow us to make one-to-one mappings between programmed values, and subproblem values, simplifying the exposition. Uniform spreading might better leverage energy scales and reduce chain breaks. Timing is not expected to be significantly impacted by acommodating the more complicated mappings, given that chain-length is two.

In addition, we make the following simplifications:
\begin{itemize}
\item[4] We omit spin-reversal transforms, these are also understood to have a negligible impact on the performance presented for cubic and Pegasus lattices and also should not be a bottleneck to timing in the C++ code when properly implemented.
\item[5] Rather than using an empirically accurate distribution of coupler and qubit defects, we simply assume the P[m=16] solver is fully yielded. Handling of a small number of defects is not expected to create a computational bottleneck.
\item[6] The QPU-API ordering for programmable couplers and qubits is chosen to take a simple form. Thus we omit the one-time cost of calculating the mapping from the linear ordering over the subproblem (variables and edges) to the linear ordering accepted by the API.
\end{itemize}

We have timed our implementation of Algorithm \ref{pseudo} using an Intel Core i7-8665U CPU @ 1.90GHz, we find the mean time required to complete 128 iterations is \SI{73.5(3)}{\milli s}. The initialization time is found to be shorter than the per iteration time, which is itself approximately half a millisecond. This timing for the cubic lattice application is significantly shorter than the reported QPU access timings, and does not qualitatively impact conclusions. Hence we justify the omission of timing information beyond the QPU-API call, as outlined in Section \ref{results}.
Our C++ code is designed to make clear the mappings and Algorithm 1 stages for didactic purposes, some further optimizations and generalizations are possible.

For toric-Pegasus and other lattices we do not analyze optimized implementations. We can anticipate that larger size and connectivity of the subproblems means a proportionately longer run time, but since the scale of Pegasus and cubic lattice subproblems are similar, we might anticipate the execution time to also be significantly shorter than QPU access time in the toric-Pegasus application also. Note further that the complications of uniform spreading and majority vote are irrelevant to the the toric-Pegasus case, since chain length is one.

\subsection{QPU parameterization}
\label{app:parameterization}

Results are presented in this paper using the DW\_2000Q\_6 and Advantage\_system\_4.1 solvers within a dwave-hybrid software framework. The code is provided open-source, and we summarise the parameterization of the QPU solver here. Non-defaulted parameters impacting QPU Access Time ($t_a$ and $n_R$) are described in Appendix \ref{app:tsQPUp}, and other parameters in Appendix \ref{app:oQPUp}.

\subsubsection{Time-sensitive QPU parameterization}
\label{app:tsQPUp}
QPU Access time is presented as a measure of latency, and is also proportional to the cost of access to a D-Wave QPU for execution of the algorithms presented. With QPU access time (\ref{eq:QPUaccess}) used as a measure of resources; it is appropriate to at least approximately tune our algorithm to minimize the energy achieved per unit of QPU access time~\cite{Raymond2020}.

In the absence of per sample and per batch overheads as part of QPU access time it might be expected that a best strategy for the hybrid method would be to draw only only one sample per iteration, particularly early in the algorithm, since a sample needn't achieve an optimum to contribute to energy minimization, and poor proposals can be rejected. For similar reasons, a relatively short anneal time could be sufficient for a proposal of sufficient quality to progress the algorithm.

In using the QPU as a subsolver we must contend with a readout overhead per sample. Assuming that we can obtain lower energy with longer anneals, in line with annealing intuitions, the use of anneal times significantly shorter than the readout time is inefficient, since energy outcomes will be poorer with no significant change in the time. Use of anneals signficantly longer than the readout time may achieve lower energies, but this is application dependent, and comes at a cost proportionate to the anneal time. For these reasons we can set the anneal duration to be comparable with the per sample overhead baseline (mostly readout time) as a default, at $t_a=$\SI{100}{\micro s}. Sampling time is then approximately equal to twice the per sample overhead basline multiplied by the number of samples.

We must also contend with a programming time, drawing additional samples and filtering for the best can only improve the quality of proposal in Algorithm \ref{pseudo}. If sampling time is much less than programming time it does not contribute much to overall time per sample added. Thus, setting sampling time significantly less than programming time is inefficient. We choose to draw $n_R=25$ samples by default, so that the programming time becomes comparable to the sampling time. These values ($t_a$ and $n_R$) are crudely tuned to work well across several D-Wave 2000Q and Advantage generation QPUs.

\begin{figure}[htb!]
  \begin{center}
    \includegraphics[width=\linewidth]{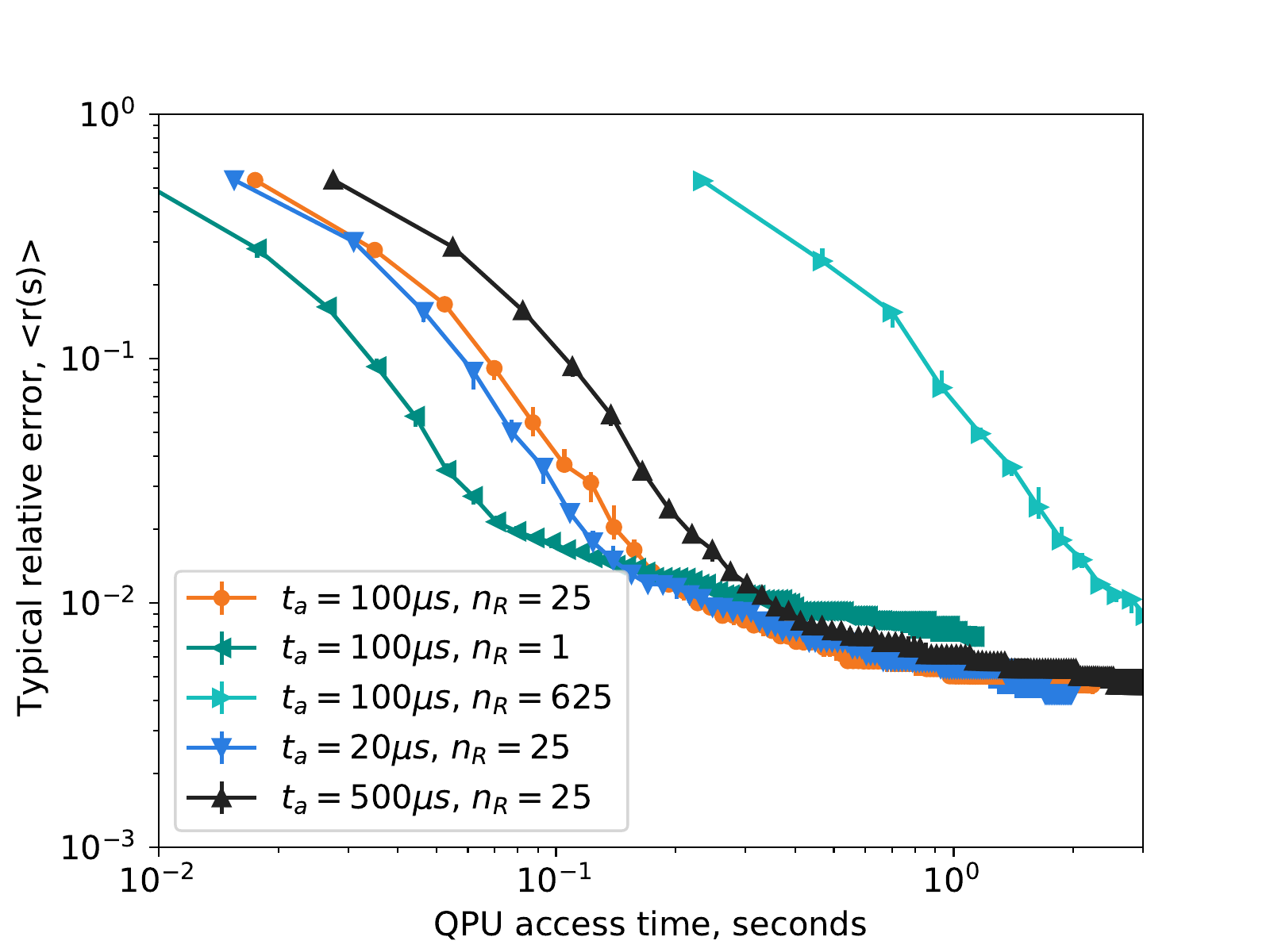}
    \includegraphics[width=\linewidth]{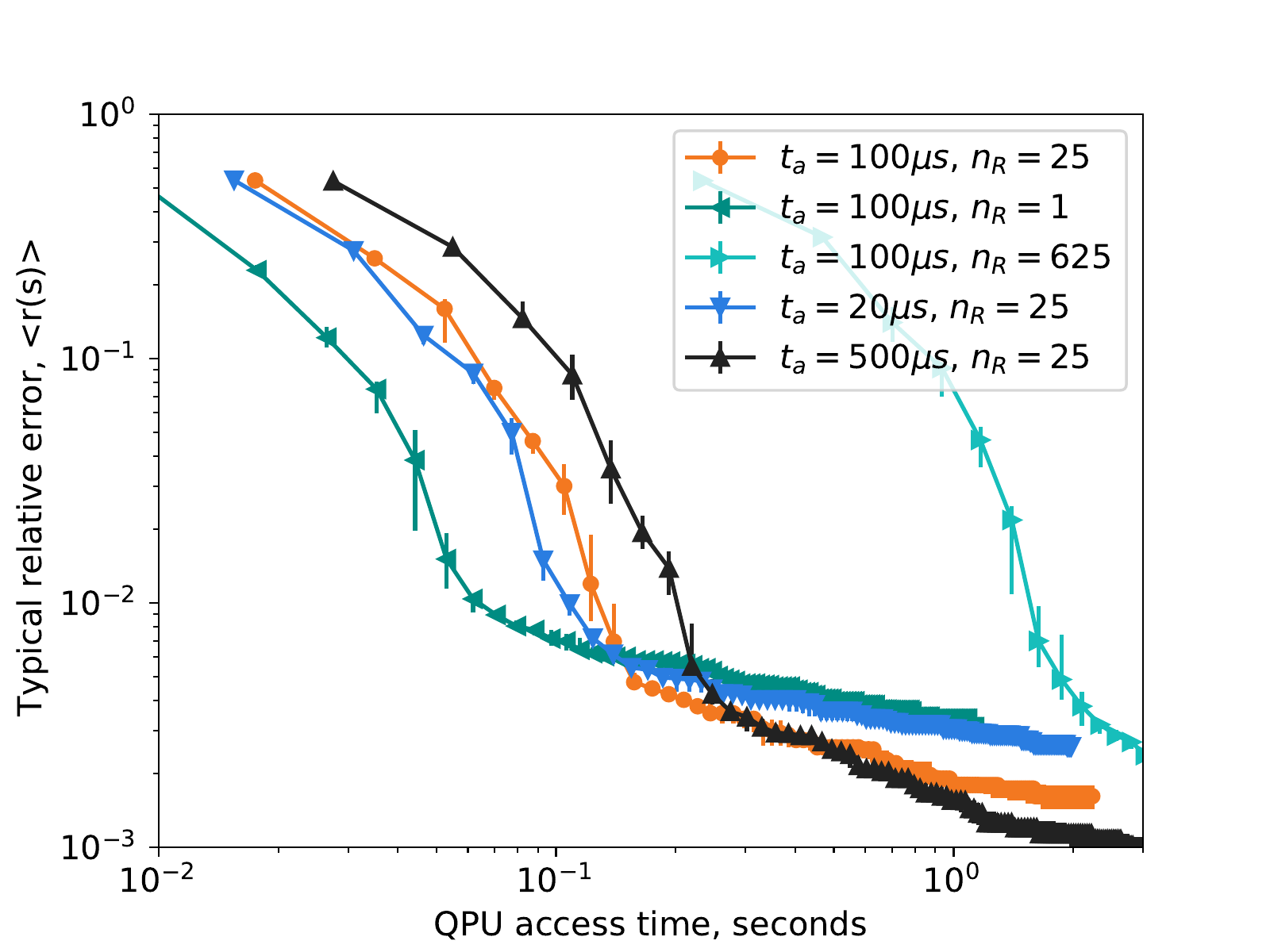}
    \caption{\label{fig:OptimizingQPUaccess} Defaults used throughout this report are an annealing time of \SI{100}{\micro s}, and 25 reads per iteration using a single programming. In this figure we examine significant variations in median relative error for periodic-cubic (top) and toric-Pegasus (bottom) spin glasses. The a-priori reasoning that we mitigate for programming and readout overheads by setting anneal time roughly equal to read out time, and programming time roughly equal to sampling time provides reasonable performance on a QPU access timescale of one second. Use of shorter anneals and fewer reads can lead to better behavior on short timescales, and use of longer anneals can be effective on long timescales.}
  \end{center}
\end{figure}

Figure \ref{fig:OptimizingQPUaccess} shows how energy achieved as a function of QPU access time is impacted by the number of samples ($n_R$) and anneal duration($t_a$). It can be seen that when we use fewer reads, or shorter anneal duration there is a gain in performance on very short timescales as expected, but relative errors are quite large through this regime. Taking larger anneal duration and/or additional reads performance per iteration improves as expected, on longer timescales this can lead to better energies.

The default is shown to performs reasonably well in both ensembles on the order of 1 second of QPU access time, but performance can clearly be improved with tuning. This tuning might also benefit by being a function of the iteration, so that cheaper (less accurate) proposals are chosen in initial iterations, progressing to higher quality more expensive proposals later on. This could be done adaptively, as a function of proposal acceptance rates for example.

\begin{figure}[htb!]
  \begin{center}
    \includegraphics[width=\linewidth]{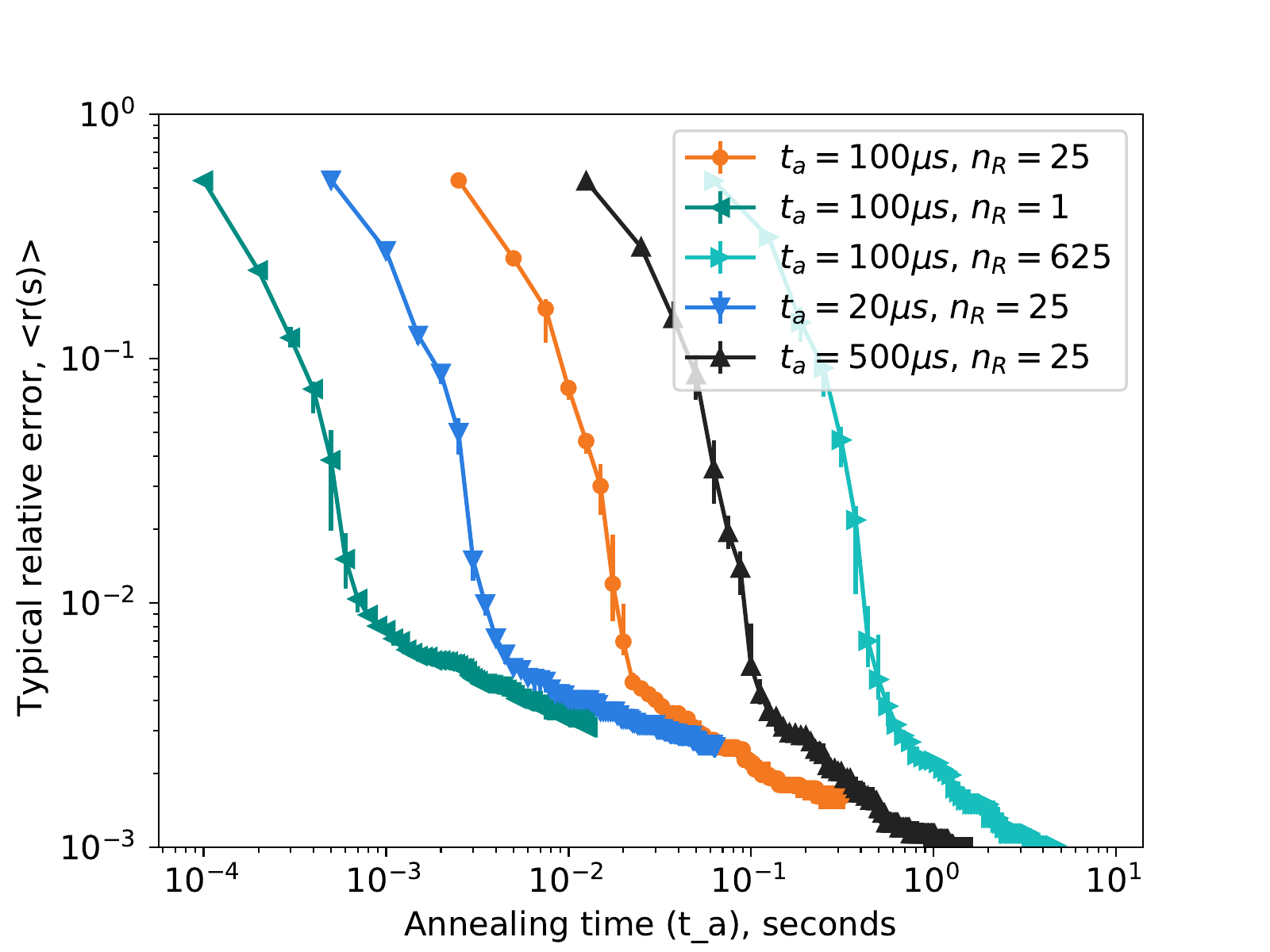}
    \caption{\label{fig:AnnealingTimeOnly} For the toric-Pegasus example of Figure \ref{fig:OptimizingQPUaccess} we can express performance as a function of annealing time only. If this is used as a measure of the algorithm time, smaller values of $n_R$ and $t_a$ are relatively more performant, compared to an evaluation as a function of QPU access time. There is also much greater variability in the timescales per iteration as a function of the parameters.}
  \end{center}
\end{figure}

It is interesting to examine the consequences of evaluating performance in terms of annealing time only, without consideration of the programming and readout overhead. Programming and readout overheads are engineering overheads that can be mitigated in principle by platform changes, whereas annealing time is a fundamental compute resource. For this reason it is common for studies to consider only annealing time, particularly for purposes of understanding scaling. Using annealing time only as shown in Figure \ref{fig:AnnealingTimeOnly} dramatically changes the picture as to algorithm efficiency, and what counts as a performant parameterization in line with the intuition provided (discounted overheads enhance the relative effectiveness at smaller $t_a$ and $n_R$). Similarly, if one were to account for additional QPU-API or algorithm overheads, we might anticipate that it might be favourable to use more compute resources per iteration (larger $t_a$ and larger $n_R$). Modification of the per QPU-API call, and per sample overheads have significant impact on the performance, and optimal parameterization, of the algorithm.

\subsubsection{Other QPU parameterization}
\label{app:oQPUp}

The toric-Pegasus application uses defaulted parameters, other than $n_R$ and $t_a$ as described in Appendix \ref{app:tsQPUp}. In the cubic lattice application we alter parameters to make use of extended J-range over chains to maximize the programmed energy scales: setting a chain strength of 2 and auto\_scale=False\cite{Harris2018,King2020}.\footnote{auto\_scale may be safely set to false as under standard embedding schemes and for the special case of the $\pm J$ ensemble examined, the programmed external fields are always within the programmable range.}

Spin reversal transforms of the subproblem (prior to embedding) are also performed for the data presented, using the SpinReversalTransformComposite~\cite{docs}. This step is omitted from Algorithm \ref{pseudo}, and from the analysis of section \ref{app:timing}. Absence of spin-reversal transforms has only a small impact on the performance presented, for the case of Pegasus lattices (where embedding is one-to-one) it can also be performed efficiently server side. 

\subsection{Sublattice vacancies and scale}
\label{app:vacancies}

\begin{figure}[htb!]
  \begin{center}
  \includegraphics[width=\linewidth]{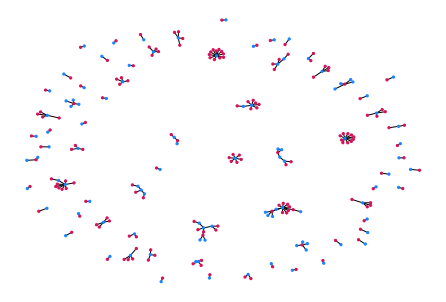}
  \caption{\label{fig:vacancies1} In order to create an origin embedding it is necessary to take special care of defects. Any chain that includes a defective qubit or coupler is omitted. The more complicated case is when two chains are coupled by an unyielded edge. In such an instance one or other (or both) chains must be added to the boundary. The pattern of variables connected by such edges in the Advantage\_system4.1 solver are shown and we must now find a selection of variables to remove, with the objective being removal of a minimal number. This is an edge-cover problem, because yield is high this can be solved optimally (the removed qubits are marked red). Patterns created for cubic lattice embeddings on DW\_2000Q\_6 and Advantage\_system4.1 also allow for brute force optimization. In cases of higher yield a heuristic could be employed.}
  \end{center}
\end{figure}

\begin{figure}[htb!]
  \begin{center}
  \includegraphics[width=\linewidth]{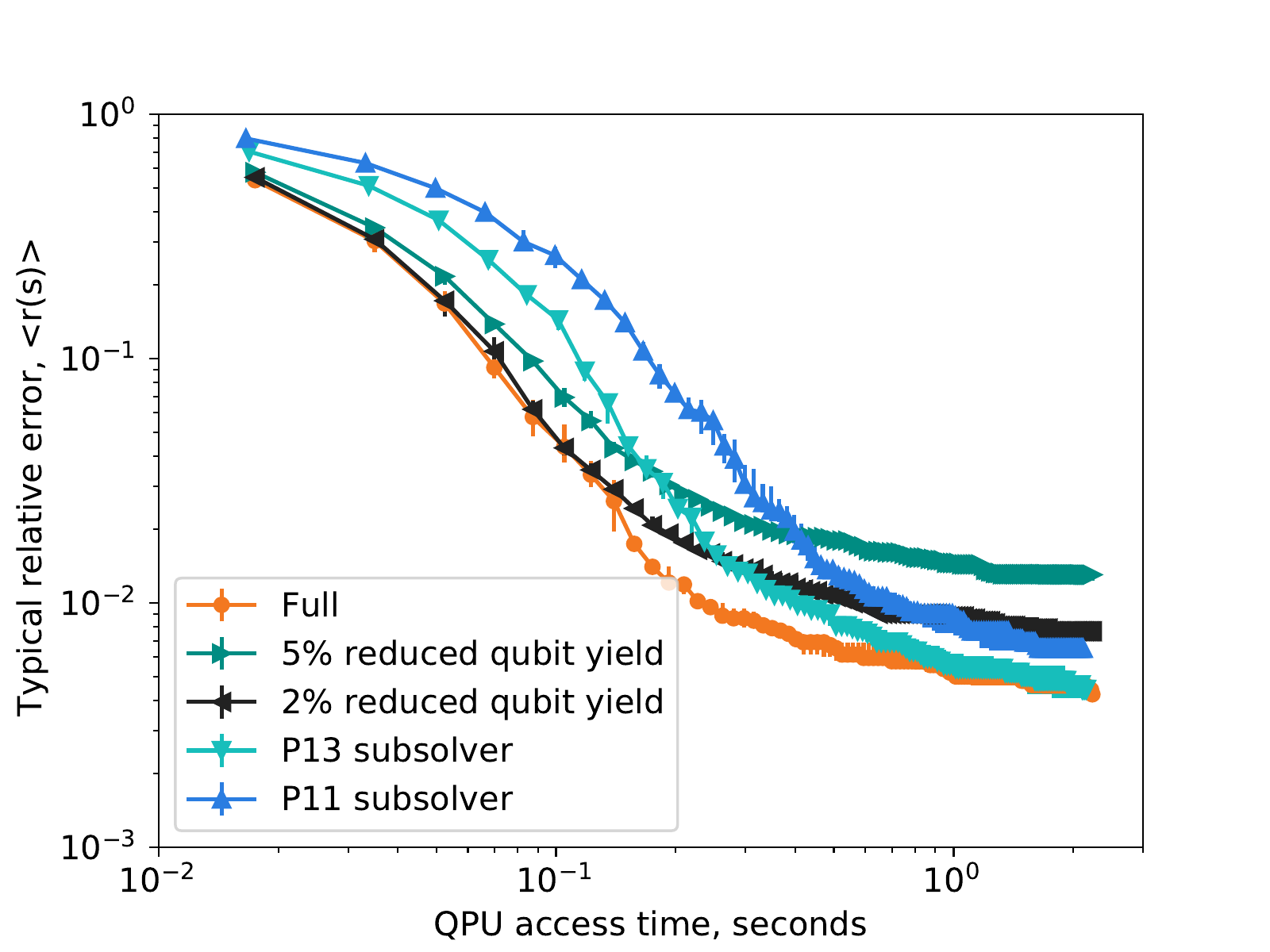}
  \caption{\label{fig:vacancies2} Cubic spin-glass median relative error with variation of the vacancy rate, or subsolver size. Vacancy rate versus optimization performance. We can artificially introduce additional defects into the Cubic lattice, so as to mimic the impact of higher rates of unyielded qubits. This can account for a significant degradataion in performance.  We can model the effect of a smaller (P13) QPU solver by similar introduction of artificial defects, this time systematically over the boundary. The solver is then able to solve 12x12x12 in place of 15x15x12. A significant degradation in performance is observed in using smaller solvers at short timescales, but the solver can catch up at longer timescales. P13 (at 100us and 25 reads) can catch up and outperform the P16 subsolver on longer timescales. As indicated by Figure \ref{fig:OptimizingQPUaccess}, it may be to get good performance at longer timescales, using the full solver, requires longer anneal durations. When size we go to smaller sizes, performance degrades quickly, consistent with results in Figures \ref{embeff1} and \ref{embeff2}.}
  \end{center}
\end{figure}

\begin{table}[h!]
  \begin{center}
    \caption{\label{table:vacancies} Qubit and coupler yield for the D-Wave Advantage and 2000Q QPUs employed in this paper. } 
    \begin{tabular}{l|l|l|l|} 
      \textbf{Subsolver name} & \textbf{Qubit yield} & \textbf{Edge yield}\\
       \hline
      Advantage\_system4.1 & 5627/5640 & 40279/40484\\
      DW\_2000Q\_6 & 2041/2048 & 5974/6016\\
    \end{tabular}
  \end{center}
\end{table}

The QPU graph can be subject to unyielded edges and couplers, these can be treated as vacancies in the regions being iterated, in effect part of the boundary to the region. Values are shown for the D-Wave 2000Q and Advanantage processors used for our analysis in Table \ref{table:vacancies}. Cases of unyielded edges that are adjacent to two yielded qubits create particularly interesting complications. As shown in Figure \ref{fig:vacancies1}, these edges may occur in a correlated manner through the graph. If an edge is unyielded one of the two qubits must be treated as part of the boundary. Eliminating the fewest variables possible becomes an edge-cover problem. Because yields are high in the DW\_2000Q\_6 and Advantage\_system4.1 we can brute force solutions, but a heuristic would likely be acceptable given lower yield.

Performance of Advantage QPUs can be evaluated as a function of artificially induced vacancies, as shown in Figure \ref{fig:vacancies2}. As vacancies are introduced we see that the burn-down is slower and that the algorithm is typically trapped by higher energy local minima on long timescales (plateauing of the relative error). Raised yields in future QPUs could raise performance. Indeed, we observed a significant performance improvement from Advantage\_system1.1 to Advantage\_system4.1, which can be explained by yield improvements.

\subsection{Workflows, parallel and post-processing}
\label{app:python}

\begin{figure}[htb!]
  \begin{center}
  \includegraphics[width=\linewidth]{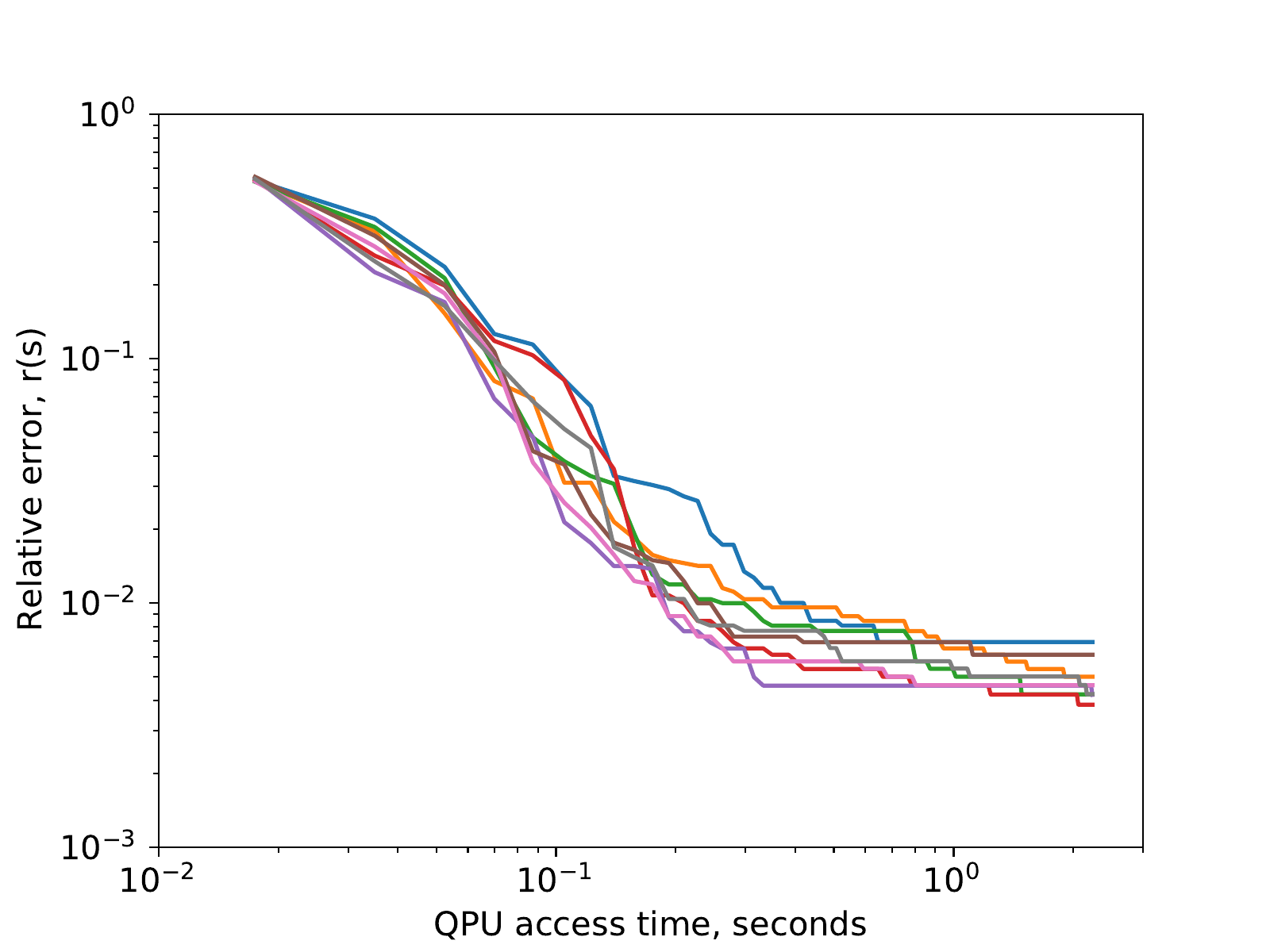}
  \includegraphics[width=\linewidth]{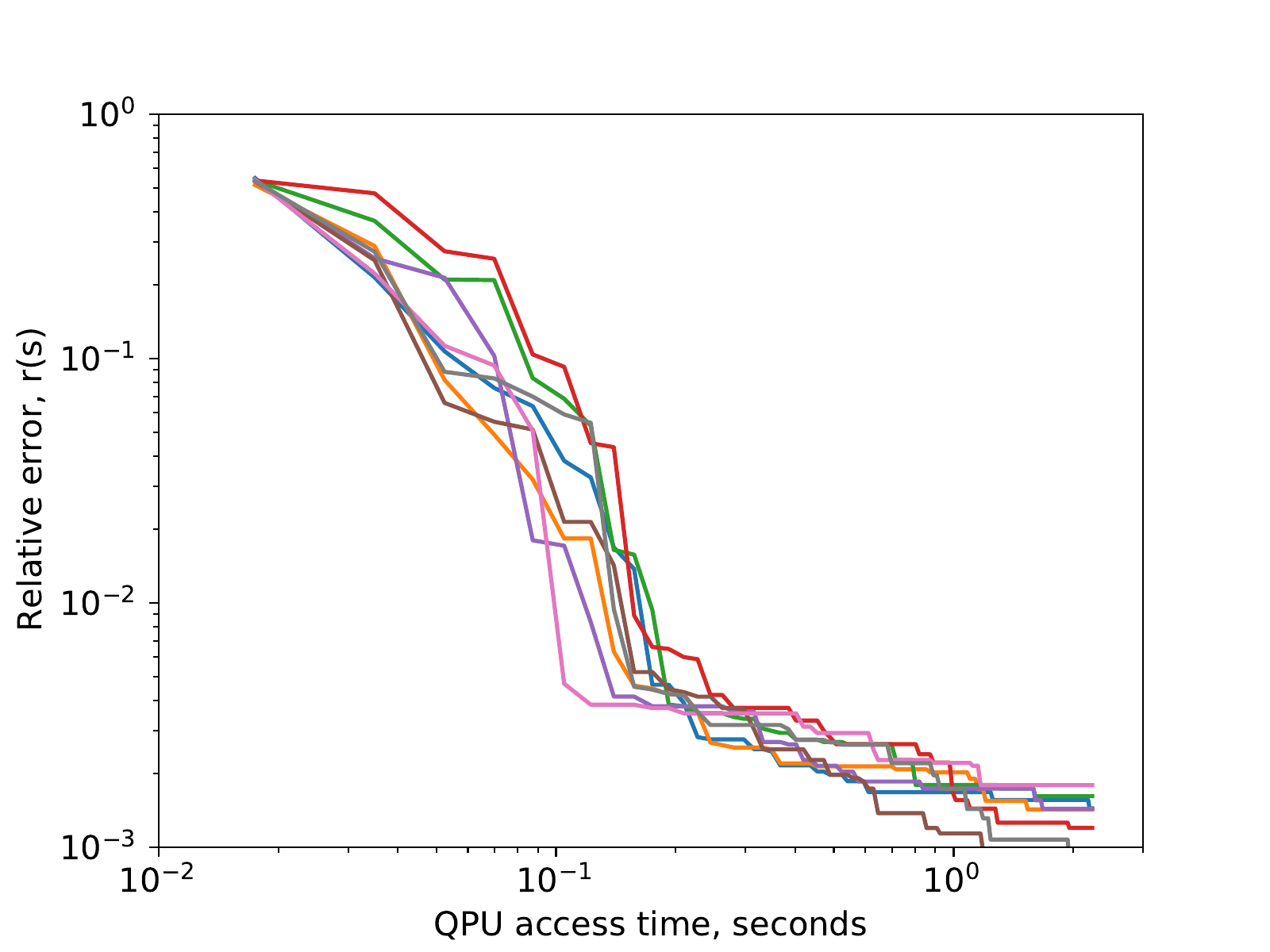}
  \caption{\label{fig:standardworkflow} Default workflow performance for 8 instances (1 run per instance) for (top) cubic periodic spin glasses and (bottom) Pegasus periodic spin glasses.}
  \end{center}
  \end{figure}

The default workflow of Algorithm \ref{pseudo} uses only the QPU, with neither parallel nor post-processing. In Figure \ref{fig:standardworkflow} statistics are shown for application of the algorithm to 8 random instances (a subset of 25 used for final statistics), with algorithmic components also randomized (the initial state, region sequence, and of course the sample sets returned by the QPU subsolver). Fluctuations from run to run are modest, particularly in the QPU access time per iteration undertaken. We choose to summarise data of this type throughout the paper by considering a median. Per iteration we take the median energy, and attribute the median QPU access time.

This workflow can be complemented with additional classical compute resources. dwave-hybrid provides straightforward means to implement such elements. In this section we examine parallel and post-processing.

While the QPU is running the local classical processor can sit idle. A means to address this is use of parallel-processing. A greedy descent implementation can be run in parallel with the QPU-Call, and typically completes first. We can consider an implementation of parallel-processing in which the current state, is subject to an optimization considering all variables, rather than as a function of the most recently solved subspace(s), though this is only one possibility.

Post-processing may improve results when the QPU differs only by a few local (perhaps thermal) excitations from the ground state~\cite{Raymond2020,Okada2019}. Greedy descent over Advantage programmable scale problems (results returned by the QPU) can be relatively fast. Greedy descent from a random initial condition requires of the order of \SI{14}{\milli s} with dwave-greedy if all the degrees of freedom available are used~\cite{Zephyr}, and greedy descent from near optima (this application) is quicker.\footnote{This value for dwave-greedy is dominated by memory copies as part of the dimod sampler interface, and may be made significantly shorter with optimization.} Thus we might consider post-processing of QPU samples as practical if we use a small number of reads, or apply post-processing only to the best proposed state modification. We consider an implementation of this post-processing applied to the subproblem, though this limitation could be relaxed.

\begin{figure}[htb!]
  \begin{center}
  \includegraphics[width=\linewidth]{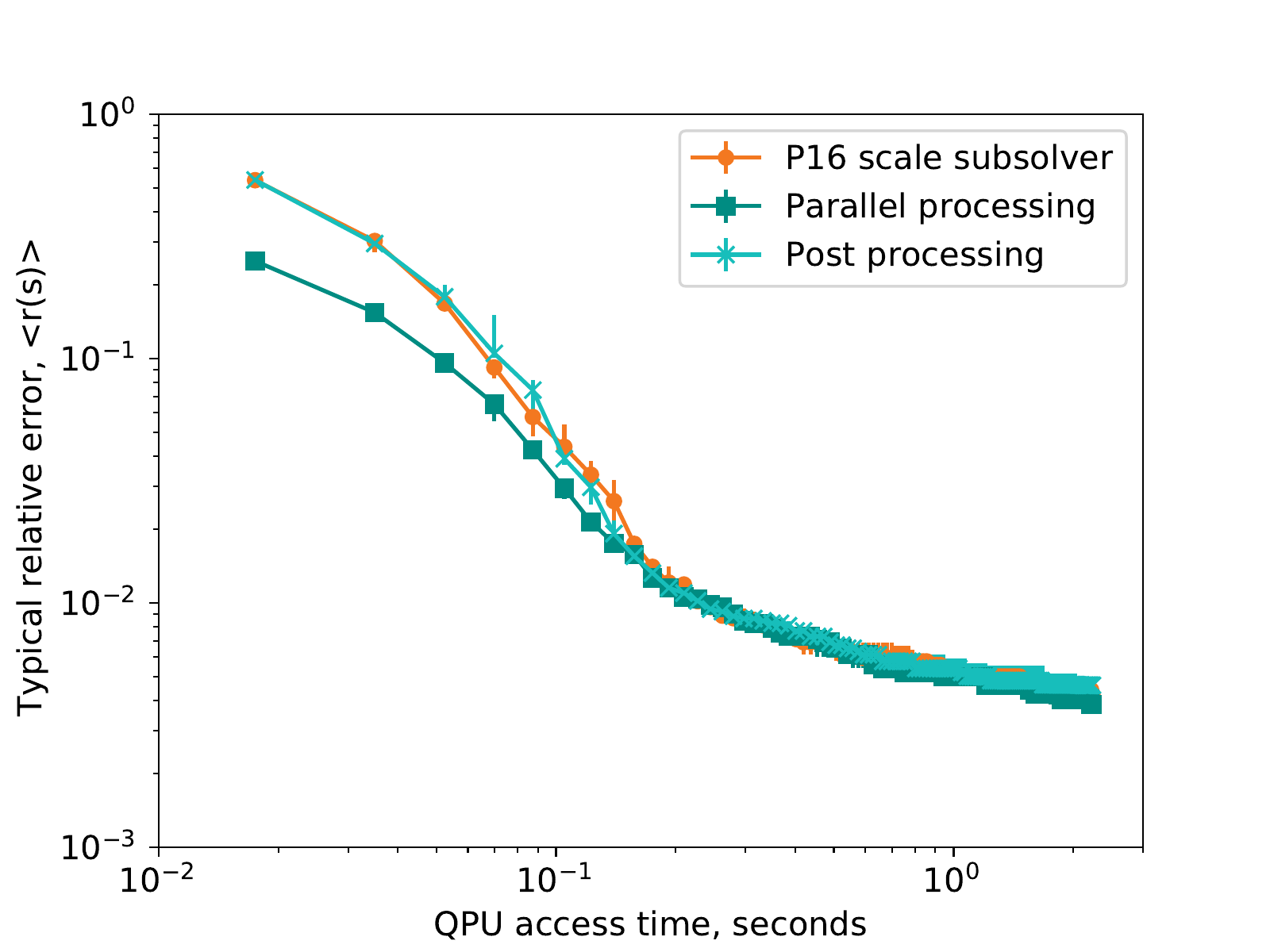}
  \includegraphics[width=\linewidth]{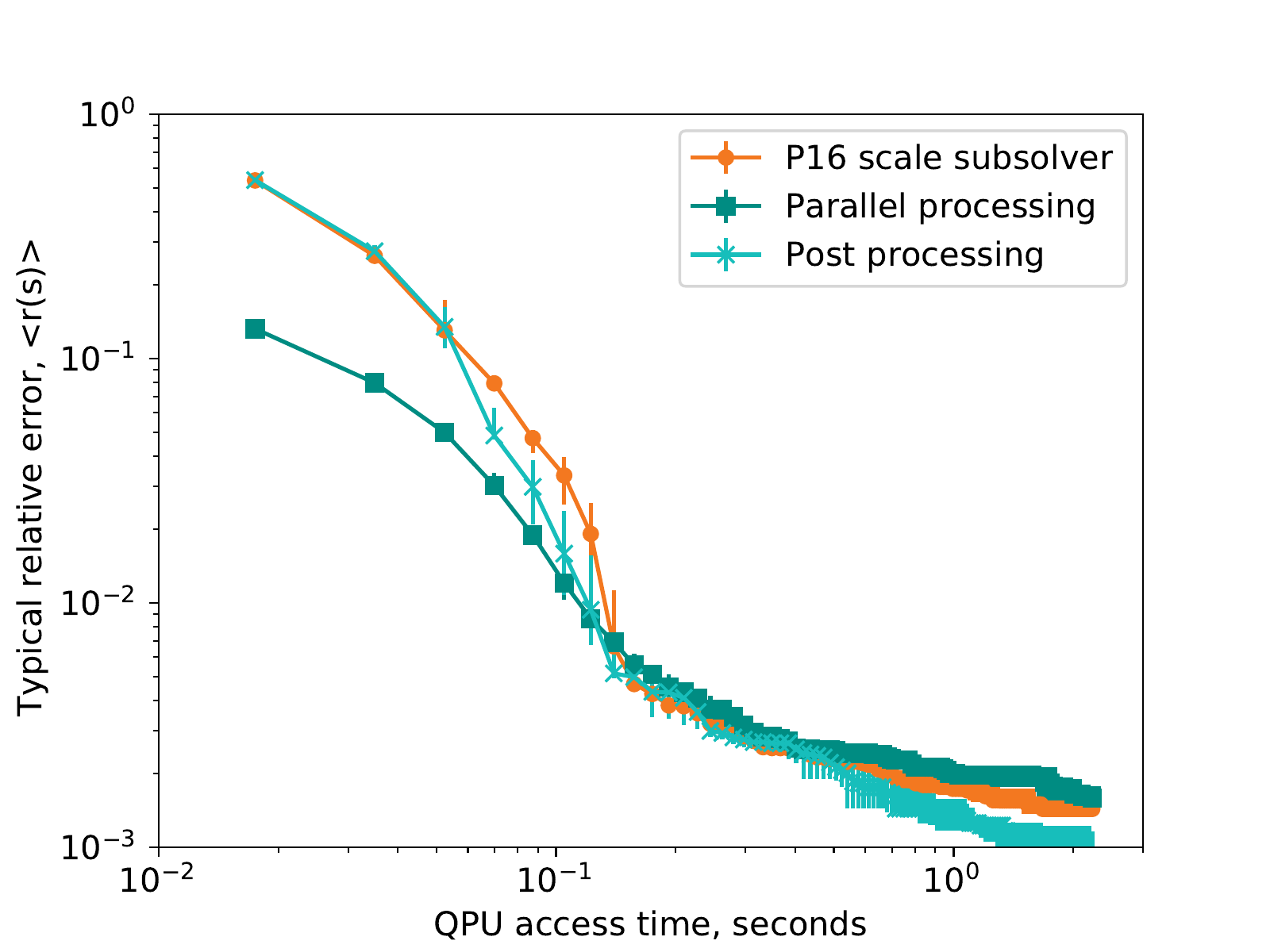}
  \caption{\label{fig:GreedyPP1} Parallel- and post-processing median relative error based on 16 problem instances are shown alongside the default workflow, for periodic-cubic (top) and toric-Pegasus (bottom) spin-glasses. At short timescales parallel-processing on the full space immediately provides a reasonable estimate over a larger set of variables, thus improving the energies returned on short time scales. Post-processing on the subspace provides an improvement in the cubic case on longer time scales.}
  \end{center}
  \end{figure}

Results for all the default workflow, along with the post-processed and parallel-processed workflows are shown in Figure \ref{fig:GreedyPP1}. Parallel-processing initially has a large impact, but on longer timescales the simpler method without parallel-processing catches up. Indeed - starting (in effect) from a local minima created by steepest greedy descent may cause enhanced trapping by local minima, and it may be advantageous to avoid this. Steepest greedy descent post-processing initially has little impact, but on longer timescales allows attainment of lower energies. On longer time scales obtaining exact optima (or, as a minimum local minima) may be important to drive down energy. The QPU might be limited by local excitations due to finite-temperature operation, and this may explain the benefits of simple (steepest-greedy) post-processing~\cite{Raymond2020}. Local excitations rates in the toric-Pegasus problem may be more common owing to smaller energy gaps and larger size, hence the greater impact of post-processing.

\clearpage

\end{document}